\documentclass[preprint]{aastex}
\usepackage{emulateapj5,apjfonts}
\usepackage{graphics}
\usepackage{amssymb}
\usepackage{bbm}
\usepackage{amsmath,textcomp,array}
\usepackage{onecolfloat}
\usepackage[ruled]{algorithm}

\usepackage{bm}
\lefthead{Child et al.}
\righthead{Halo Profiles and the $c$--$M$ Relation}

\begin{document}


\def\head{
\title{Halo Profiles and the Concentration--Mass Relation for a $\Lambda$CDM Universe}

\author{Hillary~L.~Child\altaffilmark{1,2}, Salman~Habib\altaffilmark{1,3,4}, Katrin~Heitmann\altaffilmark{1,3,4}, Nicholas~Frontiere\altaffilmark{1,2}, Hal~Finkel\altaffilmark{1,5}, Adrian~Pope\altaffilmark{1,5}, Vitali~Morozov\altaffilmark{5}}
\affil{$^1$HEP Division, Argonne National Laboratory, 9700 South Cass Avenue, Lemont, IL 60439, USA}
\affil{$^2$Department of Physics, The University of Chicago, 5640 South Ellis Avenue, Chicago, IL 60637, USA}
\affil{$^3$MCS Division, Argonne National Laboratory, 9700 South Cass Avenue, Lemont, IL 60439, USA}
\affil{$^4$Kavli Institute for Cosmological Physics, The University of Chicago, 5640 South Ellis Avenue, Chicago, IL 60637, USA}
\affil{$^5$ALCF Division, Argonne National Laboratory, 9700 South Cass Avenue, Lemont, IL 60439, USA}


\begin{abstract} 
Profiles of dark matter-dominated halos at the group and cluster scales play an important role in modern cosmology. Using results from two very large cosmological $N$-body simulations, which increase the available volume at their mass resolution by roughly two orders of magnitude, we robustly determine the halo concentration--mass $(c$--$M)$ relation over a wide range of masses, employing multiple methods of concentration measurement. We characterize individual halo profiles, as well as stacked profiles, relevant for galaxy--galaxy lensing and next-generation cluster surveys; the redshift range covered is $0\leq z \leq 4$, with a minimum halo mass of $M_{200c}\sim2\times10^{11} M_\odot$. Despite the complexity of a proper description of a halo (environmental effects, merger history, nonsphericity, relaxation state), when the mass is scaled by the nonlinear mass scale $M_\star(z)$, we find that a simple non-power-law form for the $c$--$M/M_\star$ relation provides an excellent description of our simulation results across eight decades in $M/M_{\star}$ and for $0\leq z \leq 4$. Over the mass range covered, the $c$--$M$ relation has two asymptotic forms: an approximate power law below a mass threshold $M/M_\star\sim 500-1000$, transitioning to a constant value, $c_0\sim 3$ at higher masses. The relaxed halo fraction decreases with mass, transitioning to a constant value of $\sim 0.5$ above the same mass threshold. We compare Navarro--Frenk--White (NFW) and Einasto fits to stacked profiles in narrow mass bins at different redshifts; as expected, the Einasto profile provides a better description of the simulation results. At cluster scales at low redshift, however, both NFW and Einasto profiles are in very good agreement with the simulation results, consistent with recent weak lensing observations. 

\end{abstract}

\keywords{dark matter -- galaxies: clusters: general -- gravitational lensing: weak -- methods: numerical}}

\twocolumn[\head]


\section{Introduction}
\label{sec:introduction}

The endpoints of structure formation in cold dark matter cosmologies are dark matter-dominated clumps called halos. 
In these cosmologies, initial density perturbations are amplified by the gravitational Jeans instability and small localized nonlinear structures form at high redshift. As the universe evolves, halos grow via mass accretion and halo mergers; galaxies form within halos. Halo abundance, evolution history, and properties such as mass, velocity,  sub-structure, and phase space structure, as well as the halo gas and galaxy content, all play important roles in modern cosmology, as well as in the modeling of galaxy formation, whether by empirical or semi-analytic means~\citep{white91, kauffmann93, kauffmann97,  cole94, jing98, somerville99, benson00, benson03, peacock00, seljak00, berlind02, vale04, zheng05, baugh06, conroy06, benson10, guo10, moster10, wetzel10, hearin16}. A number of observational probes based on strong and weak gravitational lensing, X-ray observations, and galaxy clustering are sensitive to the nature of halo density profiles~\citep{meneghetti05, mandelbaum06, mandelbaum08, comerford07, johnston07, okabe10, okabe13, umetsu11, umetsu14, umetsu16, oguri12, bhattacharya13, newman13, merten15, niikura15, amodeo16, okabe16, umetsu17}. This is particularly true for cluster cosmology and galaxy--galaxy lensing, which focus at the upper end of the halo mass range. 

In the remainder of the introduction, we briefly discuss halo profiles and concentrations and their importance for cosmology (Section \ref{sec:intro_profiles}), the current state of observed halo profiles (Section \ref{sec:intro_obs}), and prior work on the profiles and concentrations of simulated halos (Section \ref{sec:intro_sim}). We then outline the aims and primary results of this work (Section \ref{sec:intro_thiswork}).

\subsection{Halo Profiles}

\label{sec:intro_profiles}
Although the formation of halos is a complex, hierarchical nonlinear dynamical process, the radial density profile of individual halos is robustly fit by a surprisingly simple form, as first described by \cite{nfw1, nfw2}, using results from cosmological $N$-body simulations. The two-parameter NFW density profile is given by
\begin{equation}
\label{eqn:rho_nfw}
\rho(r) = \frac{\delta_c \rho_\textrm{c}}{\left(r/r_s\right)\left(1 + r/r_s\right)^2},
\end{equation}
where $\delta_c$ is a characteristic dimensionless density parameter. The critical density is $\rho_c(z)=3H^2(z)/8\pi G$; $H(z)$ is the Hubble parameter, and the NFW scale radius, $r_s$, is defined by the radius where the logarithmic profile slope $n_{\rm{eff}}=d\ln\rho/d\ln (r/r_s)=-2$. For $r/r_s \ll 1$, $n_{\rm{eff}}\rightarrow -1$, whereas for $r/r_s \gg 1$, $n_{\rm{eff}}\rightarrow -3$. A dimensionless shape parameter, the halo concentration, $c_{\Delta}\equiv r_{\Delta}/r_s$, is commonly used as one of the NFW parameters. The halo radius $r_{\Delta}$ is a radial scale set by the spherical overdensity (SO) halo mass definition: $M_{\Delta}\equiv (4/3)\pi r_{\Delta}^3\rho_c\Delta$, where $\Delta$ is a dimensionless overdensity parameter. We choose the critical density as the reference density; the mean density of the universe is another common choice. We also make the conventional choice of $\Delta=200$ (for X-ray work with clusters, higher values of $\Delta$ are often used, such as $\Delta=500$ or $\Delta=1000$), and refer to the corresponding concentration as $c_{200c}$.

Describing individual halos in terms of the NFW description is obviously a severe idealization. Halos are not spherical and can have complex shapes. In particular, a more realistic description of individual halos is as prolate ellipsoids with a major axis length roughly twice as long as the minor axis~\citep{jing02}. Additionally, at a fixed halo mass and more or less independent of the how the mass is defined, halo shapes and profiles can display considerable variability, with some dependence on whether the halos are dynamically relaxed~\citep{white02, lukic09}. Observations that focus on stacked halos, such as galaxy--galaxy lensing or stacked cluster weak lensing, involve averaging over many individual halos and thus reduce bias due to the characteristics of individual lenses (see, e.g., \citealt{simet17} as an example of the current state of the art). 

Despite some of these caveats, there are well-defined and observationally testable predictions for halo masses and profiles as a function of cosmological parameters. For instance, the halo mass function is an essential cosmological quantity, relevant to determining cluster abundance~\citep{holder01} and to modeling of the observed galaxy distribution, to mention two obvious examples. The halo profile shape can also be predicted accurately in modern cosmological simulations and is known to be correlated with the halo mass. One aspect of this correlation is the existence of a well-determined $c$--$M$ relation, as was already noted by NFW. Cosmological constraints delivered by ongoing surveys such as the Dark Energy Survey (DES\footnote{https://www.darkenergysurvey.org}) and next-generation surveys such as the Dark Energy Spectroscopic Instrument (DESI\footnote{http://desi.lbl.gov}), Euclid (\citealt{refregier10}), the Large Synoptic Survey Telescope (LSST) (\citealt{abell09}), and the Wide-field Infrared Survey Telescope (WFIRST) (\citealt{spergel15}) will rely on having accurate predictions for halo profiles and masses.

\subsection{Observed Profiles: Individual and Stacked Halos}
\label{sec:intro_obs}
\begin{figure}[t]
\begin{center}
 \includegraphics{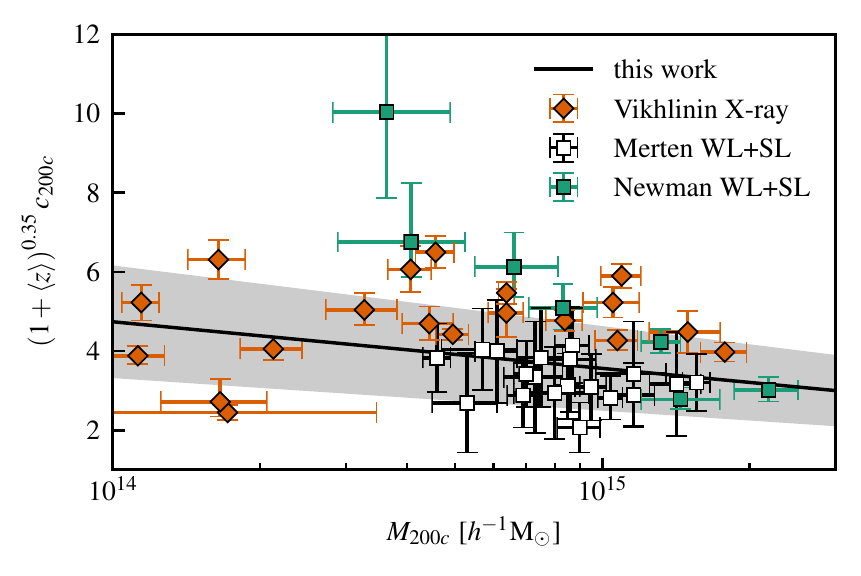}
	\caption{\label{fig:obs_indiv} Individual-halos $c$--$M$ relation with individual cluster observations using X-ray \citep{vikhlinin06, bhattacharya13} and weak and strong lensing \citep{newman13, merten15}; see Section \ref{subsec:obs} for details. The gray band represents the 1$\sigma$ intrinsic scatter in the $c$--$M$ relation, as found from the simulations.}
\end{center}
\end{figure}

Individual halo profiles can be measured using X-ray and strong and weak lensing measurements, as already mentioned. Because of a number of factors (e.g., observational limitations, selection bias, individual variability, line-of-sight dependence, analysis issues), there are difficulties in comparing these observations directly to theoretical predictions. Earlier measurements tended to have higher concentrations and a significantly steeper $c$--$M$ relation than that predicted by simulations (see, e.g., \citealt{schmidt07, broadhurst08, okabe10, oguri12} and the discussion in \citealt{comerford07}); however, the state of the art has been significantly enhanced by more recent group and cluster-scale observations. In Figure~\ref{fig:obs_indiv} we present a set of recent observational results for the $c$--$M$ relation from measurements of individual clusters. (For another compilation, see \citealt{bhattacharya13}.) We also show results from the simulation carried out in this paper, which are discussed in detail in Section~\ref{subsec:obs} below; we find good agreement between the observations and the $\Lambda$CDM predictions, despite uncertainties in accounting for selection biases and other measurement errors.

\begin{figure*}[t]
\begin{center}
	\includegraphics{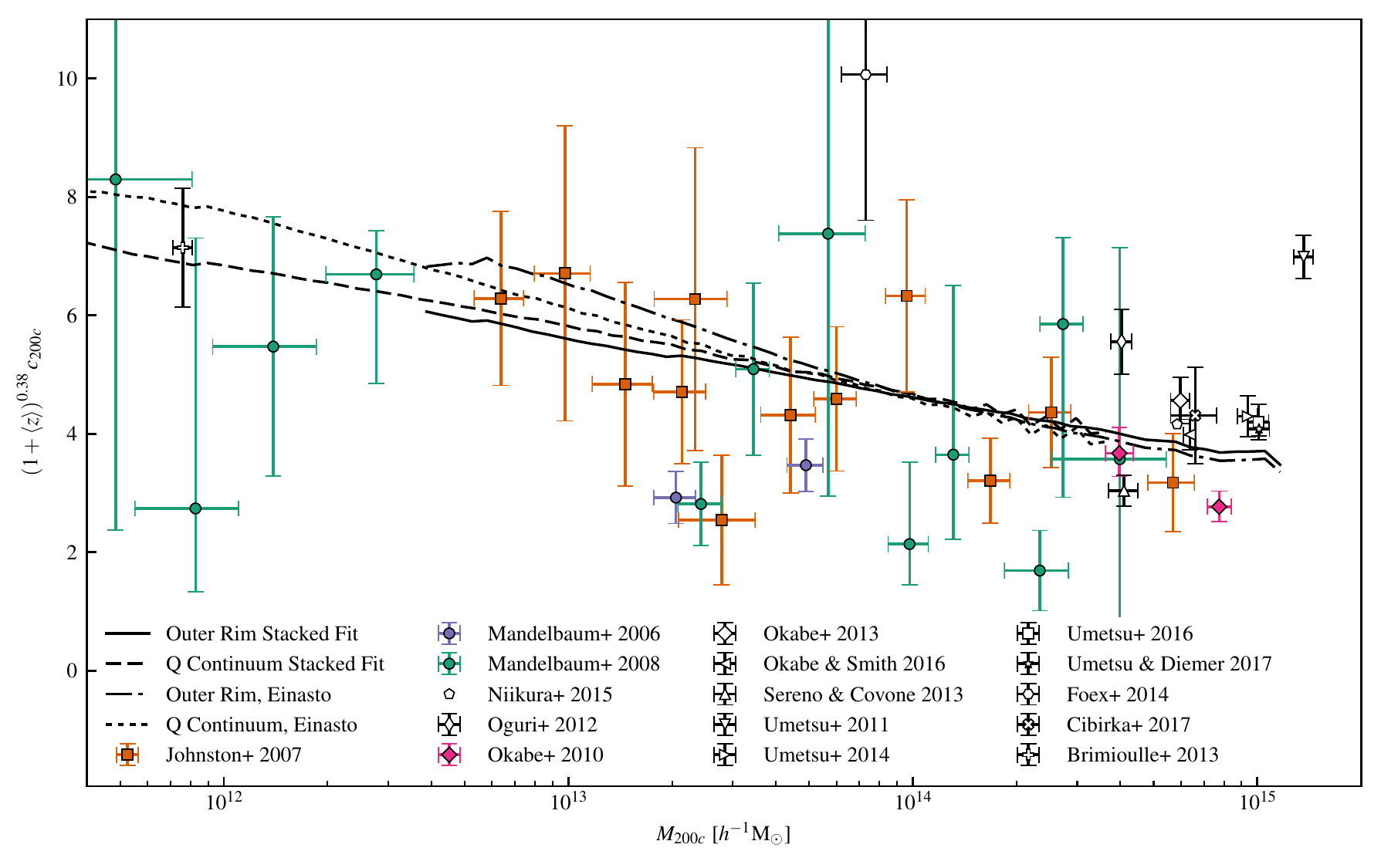}
	\caption{\label{fig:obs_ensemble} Observational results for the stacked-halo $c$--$M$ relation; see Section \ref{subsec:obs} for details. The Einasto fit yields the same concentration as NFW $c_{\rm fit}$ at high masses, but Einasto concentration rises when only one side of the peak is available to fit. Each paper represented by a  white point provides a single measurement. \cite{foex14} and \cite{umetsu11} discuss the effects of strong-lensing bias on their measurements of $c_{200c}$.}
\end{center}
\end{figure*}

Given sufficient statistics, stacking techniques can be used for both higher- (cluster weak lensing) and lower-mass halos (galaxy--galaxy lensing). The results from stacked observations average over intrinsic halo variability and lines of sight, but have different systematic issues compared to individual halo measurements. Moreover, stacked density profiles differ systematically from the NFW prescription, with potentially observable consequences. As discussed in Section~\ref{sec:stacked_halos}, the Einasto profile~\citep{einasto65} is a much better fit in this case. Next-generation surveys, and LSST in particular, will increase the number of known clusters by over an order of magnitude. Provided systematic errors can be sufficiently controlled, there is, therefore, sufficient motivation to consider the individual and stacked halo profiles separately. Figure~\ref{fig:obs_ensemble} shows observational results for stacked observations using galaxy--galaxy lensing and cluster weak lensing. Here too, the results are in good agreement; significant improvements in the observational results are expected in the near future.

\subsection{Concentrations of Simulated Halos}
\label{sec:intro_sim}
Even without including baryonic effects, which at cluster-scale masses could lead to changes at the $\sim 10\%$ level ~\citep{duffy10}, the current status of theoretical predictions and comparison with observations as shown in Section~\ref{subsec:obs} is certainly satisfactory. However, a number of open issues still remain. Simulations have shown that the concentration depends on both mass and redshift, with massive halos less concentrated than lower-mass halos at the same redshift. The $c$--$M$ relation has been measured in a large number of papers, in different ways (\citealt{bullock99}; \citealt{eke01}; \citealt{zhao03}; \citealt{duffy08}; \citealt{gao08}; \citealt{maccio08}; \citealt{klypin11, klypin16}; \citealt{ludlow12, ludlow14, ludlow16}; \citealt{prada12}; \citealt{bhattacharya13};  \citealt{correa15c}; and \citealt{diemer15}). There has been some disagreement in the results obtained, which appears to be largely due to the different ways in which the concentration has been {\em operationally} defined. Different fitting methods and binning choices can produce inconsistent $c$--$M$ relations (see, e.g., discussions in \citealt{bhattacharya13,meneghetti13,dutton14}). One of the objectives of this paper, therefore, is to present a robust set of results, arrived at by using methods with potentially different systematics, and by investigating several possible sources of numerical error.

\subsection{This Work}
\label{sec:intro_thiswork}
The primary aim of this paper is to further study and robustly determine the $c$--$M$ relation and halo profiles at group and cluster scales at low to medium redshifts, and to characterize the profile evolution over this redshift range for both relaxed and unrelaxed halos. The halo mass range considered here is mostly focused on masses significantly larger than the nonlinear (or ``collapse'') mass scale $M_\star(z)$, which gives the mass scale corresponding to peaks of the initial Gaussian random field collapsing at redshift $z$ (see definition in Section~\ref{sec:cm-rel}): $M_\star\approx 10^{12.5}\: h^{-1}M_\odot$ at $z=0$, $10^{11} \:h^{-1}M_\odot$ at $z=1$, and $10^{9.5} \: h^{-1}M_\odot$ at $z=2$. Following the discussion above, we split our $c$--$M$ relation study into two parts: one for individual halos and the other for stacked halos, in order to be consistent with current and future observational strategies.

We determine the concentration by fitting the radial mass distribution, $dM/dr$, rather than the density profile, following the procedure outlined in~\cite{bhattacharya13}. In addition to this primary method, we introduce two alternative techniques for concentration estimation -- one of which is independent of the assumed form of the density profile -- that we employ to characterize the robustness with which the $c$--$M$ relation can be determined. We find that the results are in good agreement over the mass ranges where all of our methods can be properly used. Each of the three methods is sensitive to a different range of the profile, and subject to different types of error -- we use the agreement across methods to show that our concentration measurements are robust.

The two massive state-of-the-art $\Lambda$CDM simulations (Section~\ref{sec:sims}) that form the basis of the results presented here share two essential characteristics: large volumes and very good mass resolution. At the mass resolutions considered, these simulations have roughly two orders of magnitude more volume than previous work. This allows us to robustly explore the $c$--$M$ relation with excellent statistics over a relatively wide halo mass and redshift range, and to use narrow mass bins in stacked analyses. Our results are consistent with the general idea that the concentration of halos should be set more or less by the mean density of the universe when the halos are assembled. (Unfortunately, there is no predictive theory for how the $c$--$M$ relation should depend on redshift, because there is no real theory for the NFW profile either.) At $z=0$, massive clusters, which are still forming today, would be expected to have a lower concentration than smaller-mass halos that have masses less than $M_{\star}$. Because $M_{\star}$ drops steeply with redshift, $\sim 10^{8}\: M_\odot$ at $z=3$, one might expect the $c$--$M$ relation to flatten over a significantly extended mass range as redshift increases and a large fraction of the halos in the upper mass range are still forming. Our results are very consistent with this expectation. 

By combining results from multiple redshifts and scaling the halo mass by the nonlinear mass scale $M_{\star}$, we find that the $c$--$M$ relation can be well-fit by a single expression; in agreement with~\cite{zhao03}, \cite{gao08}, \cite{ludlow14}, we find a concentration floor of $c_{200c}\sim 3$. The transition from a power-law behavior of the scaled $c$--$M$ relation to an asymptotically flat regime occurs at a halo threshold mass, $M_T\simeq 500-1000M_\star$, right at the upper end of cluster-scale halo masses at $z=0$. We note that the unrelaxed halo fraction is roughly half at masses above $M_T$, and decreases with mass below $M_T$. Further discussion of these results can be found in Section~\ref{sec:cm-rel}.  

As $M_\star$ is a function of cosmological parameters, we can compare our results to other simulations in the same family of $\Lambda$CDM cosmologies by assuming a certain level of universal behavior, as is known to hold for the mass function (see, e.g., \citealt{heitmann06}). Appendix~\ref{app:cm_others} shows that the $c$--$M/M_\star$ relations of other simulations fall within the population variance $\bar{c} \pm \bar{c}/3$ of our results, and their results from redshifts between $z=0$ and $z=3$ broadly follow the shape of a power-law transitioning to constant $c$. We have checked that this cosmology-independent behavior does not hold sufficiently far away from our fiducial cosmology by comparison to the $w$CDM results presented in~\cite{bhattacharya13} and in~\cite{kwan13}. 

The rest of this paper is organized as follows. In Section~\ref{sec:sims}, we describe the large cosmological $N$-body simulations that are the source of the halo catalogs and halo profiles. Section~\ref{sec:conc} describes our methodology for measuring the halo concentrations using the radial profiles. Section~\ref{sec:cm-rel} presents the resulting $c$--$M$ relations and the new fitting form using scaled halo masses. Section~\ref{sec:disc} concludes with a final discussion of the results. The appendices contains the results of investigations of possible sources of numerical error and comparisons of some of our results to previous work.

\section{Simulations}
\label{sec:sims}
The results reported here use data from two very large, gravity-only $N$-body simulations run with the Hardware/Hybrid Accelerated Cosmology Code (HACC) framework \citep{habib14}. These are the `Q Continuum'~\citep{QCont} and `Outer Rim'~\citep{habib14} simulations carried out on the CPU/GPU system Titan at Oak Ridge National Laboratory and the Blue Gene/Q (BG/Q) system Mira at Argonne National Laboratory, respectively. HACC uses a hybrid force calculation scheme, splitting the total force calculation into a long-range component and a short-range component. In both runs, the long-range forces are computed using the same high-order spectral particle mesh method, while the short-range forces are computed using different methods (albeit with the same hand-over scale to the short-range solver) in order to best exploit the available computational architecture. A direct particle-particle interaction technique is used for the CPU/GPU system and an RCB (Recursive Coordinate Bisection) tree method for the BG/Q system. Halo identification and characterization is carried out with HACC's parallel CosmoTools analysis framework, using a combination of in situ and offline analyses. 

The Q~Continuum and Outer~Rim runs represent independent realizations of the same shared WMAP-7 \citep{komatsu11} cosmology:
\begin{eqnarray}
\omega_{\rm cdm}&=&0.1109\stackrel{h=0.71}{\Rightarrow}\Omega_{\rm
  cdm}=0.220,\nonumber\\ 
\omega_{\rm b}&=&0.02258,\nonumber\\
n_s&=&0.963,\nonumber\\
h&=&0.71,\nonumber\\
\sigma_8&=&0.8,\nonumber\\
w&=&-1.0,\nonumber\\
\Omega_{\nu}&=&0.0,
\end{eqnarray}
but with differing volumes and mass resolution. The box size for the Q Continuum run is $L_{\rm QC}=1300~{\rm Mpc}=923 \: h^{-1}\:{\rm
  Mpc}$, while that of Outer Rim is $L_{\rm OR}=4225~{\rm Mpc}=3000 \: h^{-1}\:{\rm Mpc}$. The number of particles in these simulations are $8192^3=0.55$~trillion (Q Continuum) and $10240^3=1.1$~trillion (Outer Rim); the associated mass resolutions are $m_p=1.48\times 10^{8}~M_\odot=1.05\times 10^8 \: h^{-1}M_\odot$ (Q Continuum) and $m_p=2.6\times 10^{9}~M_\odot=1.85\times 10^9 \: h^{-1}M_\odot$ (Outer Rim). The force resolutions are (comoving) $2\:h^{-1}\: {\rm kpc}$ (Q Continuum) and $3\:h^{-1}\: {\rm kpc}$ (Outer Rim). Both simulations are given a Zel'dovich approximation initial condition at $z=200$ with transfer functions generated by the CAMB code~\citep{lewis00}. The Outer~Rim simulation has been used for several analyses of SDSS IV extended Baryon Oscillation Spectroscopic Survey data \citep{gilmarin18, hou18, zarrouk18}.

The large volumes and excellent mass resolution in these simulations lead to the following advantages in characterizing halo properties: (1) sufficiently large numbers of halos at high masses over the redshift ranges studied (at $z=0$, $\sim 20$ million and $\sim 10$ million halos of at least 2000 particles in Q Continuum and Outer Rim, respectively); (2) excellent profile resolution for individual halos; and (3) the ability to study stacked halo profiles in narrow mass bins -- hundreds of halos in mass bins of width $ \pm 5\%$ at cluster scales, and hundreds of thousands at lower masses. Compared to our previous work in~\cite{bhattacharya13}, the mass resolution is improved by more than an order of magnitude. In addition, the overlapping volume and mass resolution coverage between the two boxes (which are run using different $N$-body algorithms) provides an automatic cross-check for certain types of systematic errors that can arise in cosmological simulations.

\section{Concentration Measurement}
\label{sec:conc}

In this section, we describe our methods for measuring halo concentrations. This requires first defining and measuring the halo mass, followed by a determination of the halo concentration. In the context of the $c$--$M$ relation, the mass is usually defined in terms of an SO, $\Delta$, as discussed in Section~\ref{sec:introduction}. A halo with mass $M_{\Delta}$ has a corresponding size, $r_{\Delta}$, the radius within which the halo has an average overdensity of $\Delta$ with respect to $\rho_{c}$. Common choices of $\Delta$ include $\Delta=200$ and $\Delta=\Delta_{{\rm vir}}$, where $\Delta_{{\rm vir}}$ follows from the spherical top-hat collapse model. It is also not uncommon to define the overdensity with respect to the mean density of the universe, rather than the critical density. As stated earlier, we will use the critical density as the reference, with $\Delta=200$.
It should be noted that the definition of the SO mass does not depend on the nature of the density profile. 

A common alternative NFW parameterization, which we use here, describes the NFW profile in terms of an SO halo mass and the halo concentration, $c_{\Delta}$. Written in terms of the SO radius $r_\Delta$ and the concentration, the NFW profile becomes
\begin{equation}
\rho(r)=\frac{\Delta\rho_c}{3A(c_{\Delta})}
\frac{1}{(r/r_{\Delta})(1/c_{\Delta}+r/r_{\Delta})^2}, 
\label{nfw2}
\end{equation}
where $A(c_{\Delta})\equiv\ln(1+c_{\Delta})-c_{\Delta}/(1+c_{\Delta})$. 

Using the NFW profile as defined by Equation~\ref{nfw2}, it is clear that given $r_{\Delta}$ (or equivalently, $M_{\Delta}$) and $c_{\Delta}$ (or equivalently, $r_s$), the NFW profile is uniquely determined. Given the NFW description, the SO mass and concentration together completely determine the spherically averaged halo profile. 

Halo concentrations are computed using halo profiles built by HACC's parallel SO halo finder, which is part of the CosmoTools analysis framework. First, friends-of-friends (FOF) halos with dimensionless linking length $b=0.168$ are found by a fast, parallel, tree-based algorithm, and their centers determined by finding the deepest potential minimum within the FOF halo. All particles (not just those in the original FOF halo) are counted in radial shells centered on the point of minimum potential, and the mass $M_{200c}$ is calculated as the mass within a sphere whose average density is 200 times the critical density. No unbound particles are removed because we are interested in the density profile as measured and not in some idealized theoretical notion of what might constitute membership in a halo. Twenty shells are placed uniformly in log space between a minimum radius at the smoothing scale and a maximum radius greater than $R_{200c}$, and the differential mass profile $dM/dr$ is calculated from the bin widths and particle counts. We note that the notion of an SO halo becomes problematic during major halo mergers; the halo center is also potentially not well-defined during such epochs.

The concentration measurement procedure uses three different methods, explained in detail below: profile fit (as in \citealt{bhattacharya13}), accumulated mass, and peak finding. The idea behind using these different methods is to explore the robustness with which the concentration can be determined in the presence of different types of systematic errors. We measure concentrations only for well-sampled halos, i.e., those above a conservative threshold of at least 2000 particles within $r_{200c}$ (see Appendix \ref{sec:min_particle_count}), corresponding to a mass of $2.1 \times 10^{11} \: h^{-1} M_\odot$ for Q Continuum and $3.7 \times 10^{12} \: h^{-1} M_\odot$ for Outer Rim. 

Following standard practice, for all three methods we keep the mass $M_{200c}$ fixed as found by the SO algorithm described above and fit only for the concentration. In principle, it is possible to allow both quantities to float, but this leads to variability in the determined concentrations, even if the associated $M_{200c}$ changes only by a small amount.

\subsection{Individual Halos}
\label{sec:ind_halos}
\begin{figure}[htb]
	\centerline{
    \includegraphics{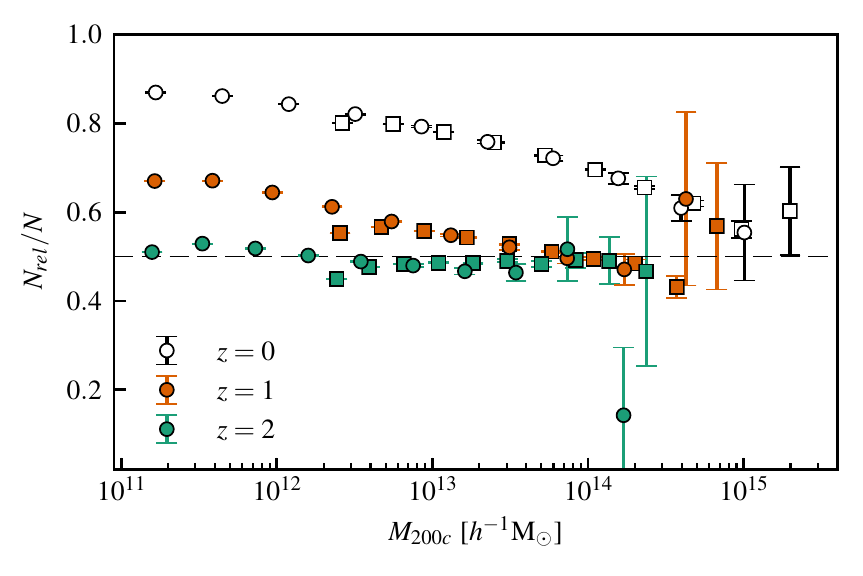}}
	\caption{Relaxed fraction for Q~Continuum (circles) and Outer Rim (squares) halos. A halo is considered relaxed if the distance between its most bound particle and SO center of mass is, at most, $0.07R_{200}$.}
\label{fig:rel_frac_m}
\end{figure}

Halos are dynamically evolving objects; a halo profile may not be well-described by the NFW profile if it is far from a dynamically relaxed state -- for example, if it is the product of a recent merger. We identify these halos by a simple test~\citep{neto07, duffy08}: halos are labeled relaxed if the distance between the halo center and the center of mass of all particles in the SO halo is, at most, $0.07R_{200c}$. If the offset exceeds $0.07R_{200c}$, the halo is assumed to be unrelaxed. (In rare circumstances, characterized by accidental symmetry, unrelaxed halos can pass the relaxed halo test, but the reverse is not true.) At $z=0$, $\sim 80\%$ of all halos of at least 2000 particles are relaxed, with a higher fraction of high-mass halos unrelaxed. As redshift increases, the relaxed fraction drops to 45\%, and by $z=2$ is independent of mass across the $10^{11}\:h^{-1} {M}_\odot < M_{200c} < 10^{15}\:h^{-1}{ M}_\odot$ range (see Figure \ref{fig:rel_frac_m}). 

\begin{figure*}[htb]
	\centerline{
   	\includegraphics{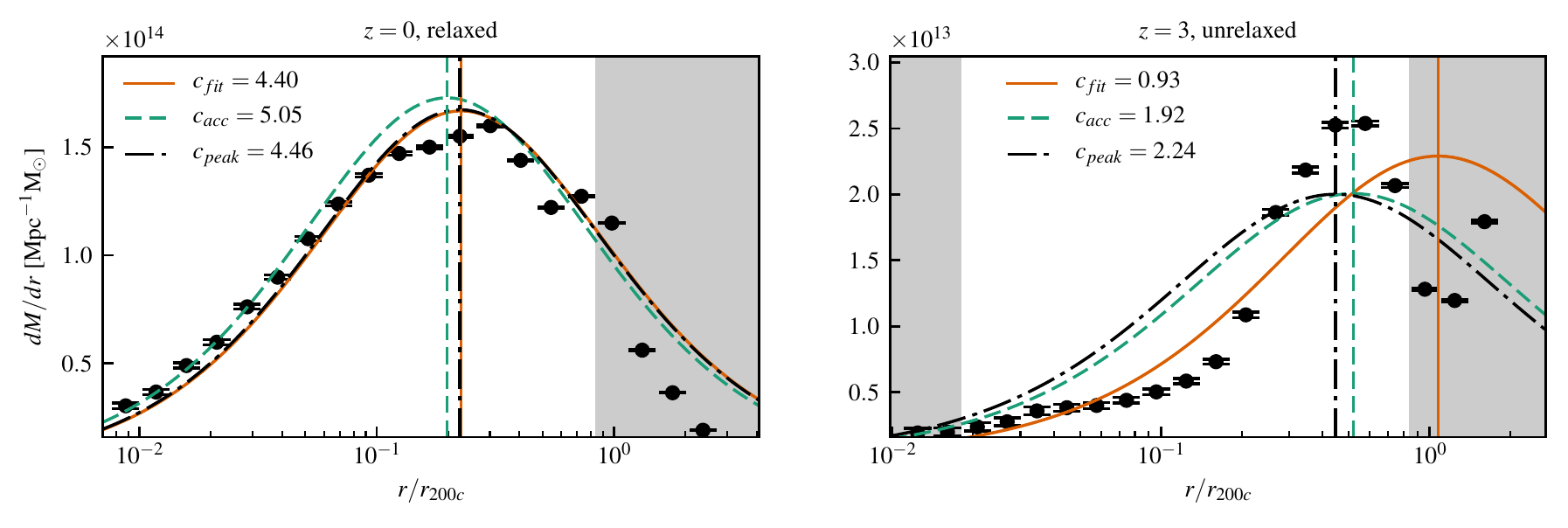}}
	\caption{\label{fig:rep_indiv_halos} Example profiles of individual high-mass Q~Continuum halos well- and poorly fit to an NFW profile: a relaxed halo at $z=0$ (left, $M_{200c} = 10^{14} \: h^{-1}{M}_\odot$) and an unrelaxed halo at $z=3$ (right, $M_{200c} = 9 \times 10^{12} \: h^{-1}{M}_\odot$). Note the small Poisson error. At $z=0$ the outer radius of the innermost bin encloses at least 100 particles, so all points within $r_{200c}$ are included in the profile fit. At $z=3$, the first two bins do not meet this criterion and are dropped. More points must be dropped for lower-mass and less-concentrated halos. Shaded regions are not included in the fit, nor expected to follow an NFW profile.}
\end{figure*}

\begin{figure}[htb]
	\centerline{
   	\includegraphics{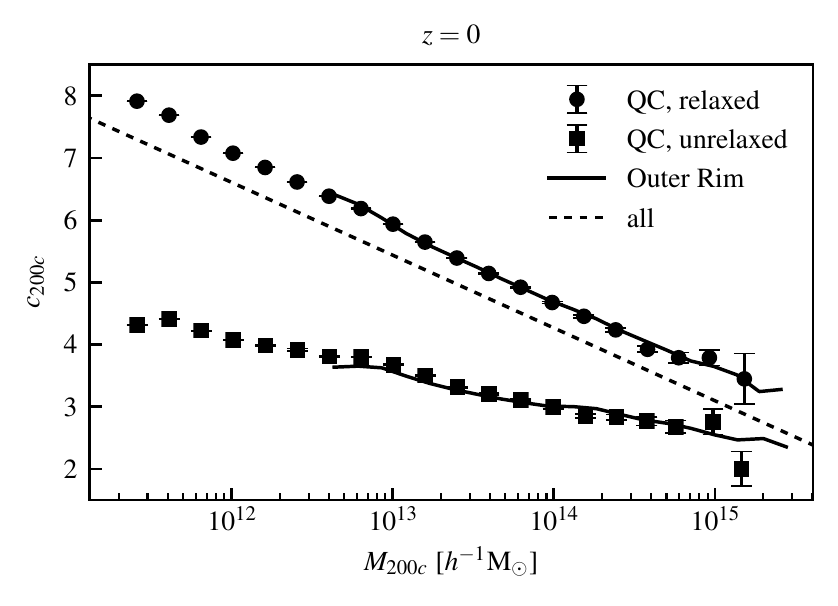}}
	\caption{Mean fit concentrations at $z=0$ for all relaxed and unrelaxed halos from the Outer Rim and Q~Continuum (QC) runs using the profile fit method. At this redshift, 80\% of halos of at least 2000 particles are relaxed. The dotted line shows the power-law fit to all halos; the relaxed halos have slightly higher concentration (upper data set), while the $c$--$M$ relation for unrelaxed halos is lower and flatter (lower data set). Note the excellent agreement between the two simulation results.}
\label{fig:rel_unrel}
\end{figure}

Figure \ref{fig:rep_indiv_halos} shows examples of relaxed and unrelaxed halo profiles at high and low redshifts. Note that much of the mass of the unrelaxed halo is far from the center, so the identified scale radius is large and concentration is small. In general, our relaxation criterion implies that unrelaxed halos have significant mass far from the potential minimum, so we expect and find unrelaxed halos to be less concentrated than relaxed halos of the same mass (Figure \ref{fig:rel_unrel}). The unrelaxed halo $c$--$M$ relation has lower amplitude and slope than the $c$--$M$ relation of all halos, but as for relaxed halos, concentration decreases with mass at low redshift and is constant at high $z$. Notably, this holds true at high masses and redshifts -- the relaxed halo $c$--$M$ relation does not fall below the $c$--$M$ relation for unrelaxed halos. 

Individual halos are fit to the NFW differential mass profile,
\begin{equation}
\frac{dM}{dr} = 4\pi r^2\rho(r)=\frac{M_\Delta}{A(c_\Delta)\:r_\Delta} \frac{r/r_\Delta}{(1/c_\Delta + r/r_\Delta)},
\label{dmdr}
\end{equation}
which rises as $r$ at small $r$, peaks at $r=r_s$, and falls off as $1/r$ using the following three methods: 

\textit{Profile Fit, $c_{\rm fit}$.} The halo mass $M_\Delta$ and radius $r_\Delta$ are fixed by $M_{200c}$ as found by the SO algorithm. The profile fit uses the Levenberg--Marquardt algorithm, which, weighted by the Poisson error in the number of particles in each shell, minimizes
\begin{equation}
\sum_i \frac{\left\lbrack dn_i/dr_i - (dn/dr)_i^\textrm{NFW} \right\rbrack^2}{dn_i / dr_i^2},
\end{equation}
where $dr_i$ is the radial width of a shell that contains $dn_i$ particles and $(dn/dr)_i^\textrm{NFW}$ is evaluated at the midpoint of the bin. Only shells whose outer radius falls within $r_{200}$ and encloses at least 100 particles are fit. Shells beyond $r_{200}$, with their high particle counts and low Poisson error, would have disproportionate influence on the fit, and the NFW form does not necessarily hold at the farther edges of a halo. The requirement of at least 100 particles in a shell also excludes the inner regions (roughly a tenth of the virial radius at cluster mass scales) that may suffer from numerical errors and not be modeled well by gravity-only simulations (due to missing baryonic/feedback effects).

\textit{Accumulated Mass, $c_{\rm acc}$.} This method uses the fact that the mass enclosed by the NFW scale radius is
\begin{equation}
\label{eqn:accmass}
M(r_s) = \frac{M_\Delta}{A(c_\Delta)}\left(\textrm{ln} \:2 - \frac{1}{2} \right).
\end{equation}
The concentration is found iteratively by fixing $c_\Delta$, interpolating the enclosed mass profile to solve Equation~(\ref{eqn:accmass}) for $r_s$, and updating $c_\Delta = r_\Delta/r_s$. 

\textit{Peak Finding, $c_{\rm peak}$.} The differential mass profile, Equation~(\ref{dmdr}), peaks at $r = r_s$, so the scale radius can be measured by simply locating the peak. To do this, profiles are smoothed using a three-point Hanning filter
\begin{equation}
f(r_i) = \frac{1}{4}\left[f(r_{i-1})+ 2f(r_i) + f(r_{i+1})\right],
\end{equation}
and the scale radius is set to the location of maximum smoothed $dM/dr$, excluding the first and last radial bins. Note that this method makes no assumption about the specific form (NFW, Einasto, etc.) of the halo profile.

Our goal in using three different methods is to verify the robustness of our concentration measurements. These three methods are sensitive to different features of the profile. Both the profile fit and accumulated mass methods assume the NFW form, while peak finding does not. The accumulated mass method counts particles only to the scale radius, so it is most sensitive to the inner profile, $r<r_s$. The profile fit is more influenced by the outer profile, as the outer shells, with their higher particle counts, have lower Poisson error and are weighted more heavily in the fit. If halos were always well-described by the NFW form, all three methods would find the same concentration; when their results differ, the profile is not a perfect NFW profile. We use $c_{\rm fit}$ as the primary concentration measurement and the other two methods to check our results and to better understand changes in halo profiles with mass and redshift.

\begin{figure*}[t]
\begin{center}
	\includegraphics{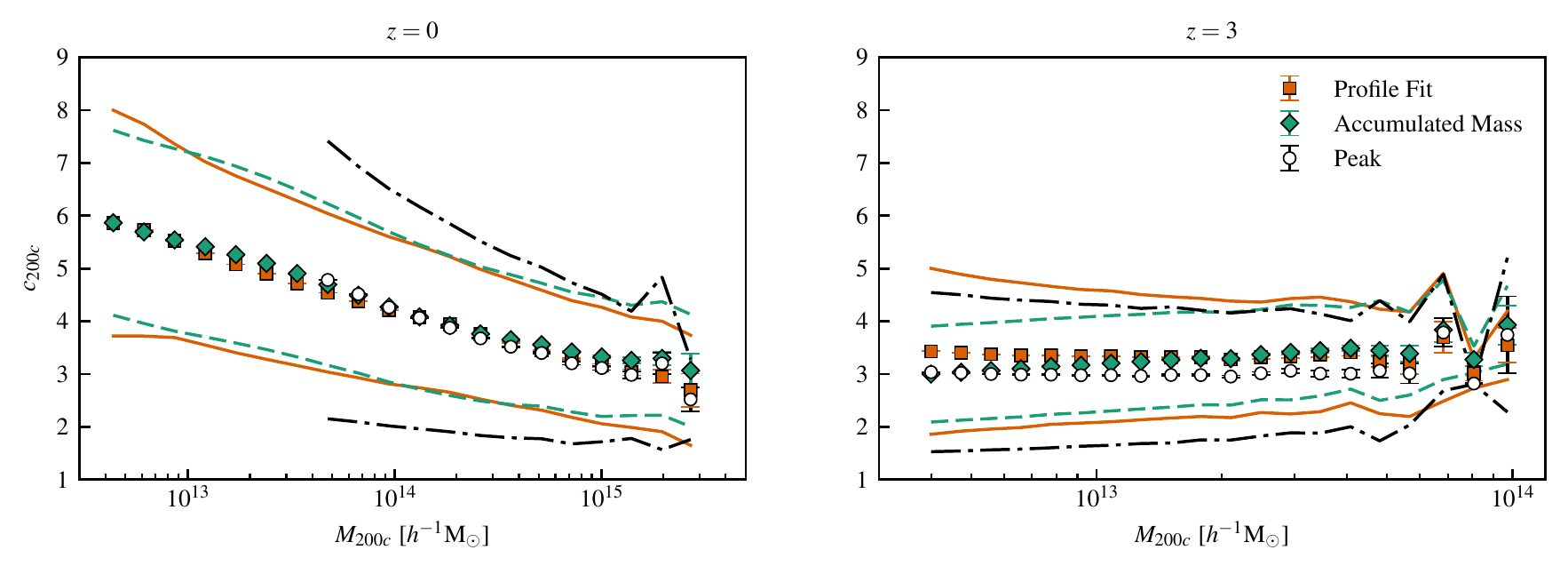}
	\caption{Profile fit, accumulated mass, and peak finding methods of concentration measurement for all Outer Rim halos at $z=0$ (left panel) and $z=3$ (right). The outer curves represent the $1\sigma$ intrinsic variation in the concentration at a fixed mass. Note the very small statistical error bars -- the lowest and highest mass bins shown contain $10^6$ and 11 halos, respectively, at $z=0$, and $10^5$ and 4 halos at $z=3$. Here, $c_{\rm peak}$ is shown only for high-mass, low-concentration halos; see further discussion in Appendix \ref{app:peak_method}. For high-mass halos at $z=0$, there is little difference in mean concentration between methods; at $z=3$, $c_{\rm peak}$ differs from $c_{\rm fit}$ by about 10\%.}
\label{fig:methods1} 
\end{center}
\end{figure*}

The three methods agree best on well-resolved halos at low redshift, as shown in Figure \ref{fig:methods1}. At high redshift, the methods differ by as much as 10\%: the fit and peak concentrations are essentially flat as a function of mass, while the accumulated mass $c$--$M$ relation slopes slightly upward, still well within the range of the intrinsic concentration scatter.

\begin{figure*}[t]
	\centerline{
   	\includegraphics{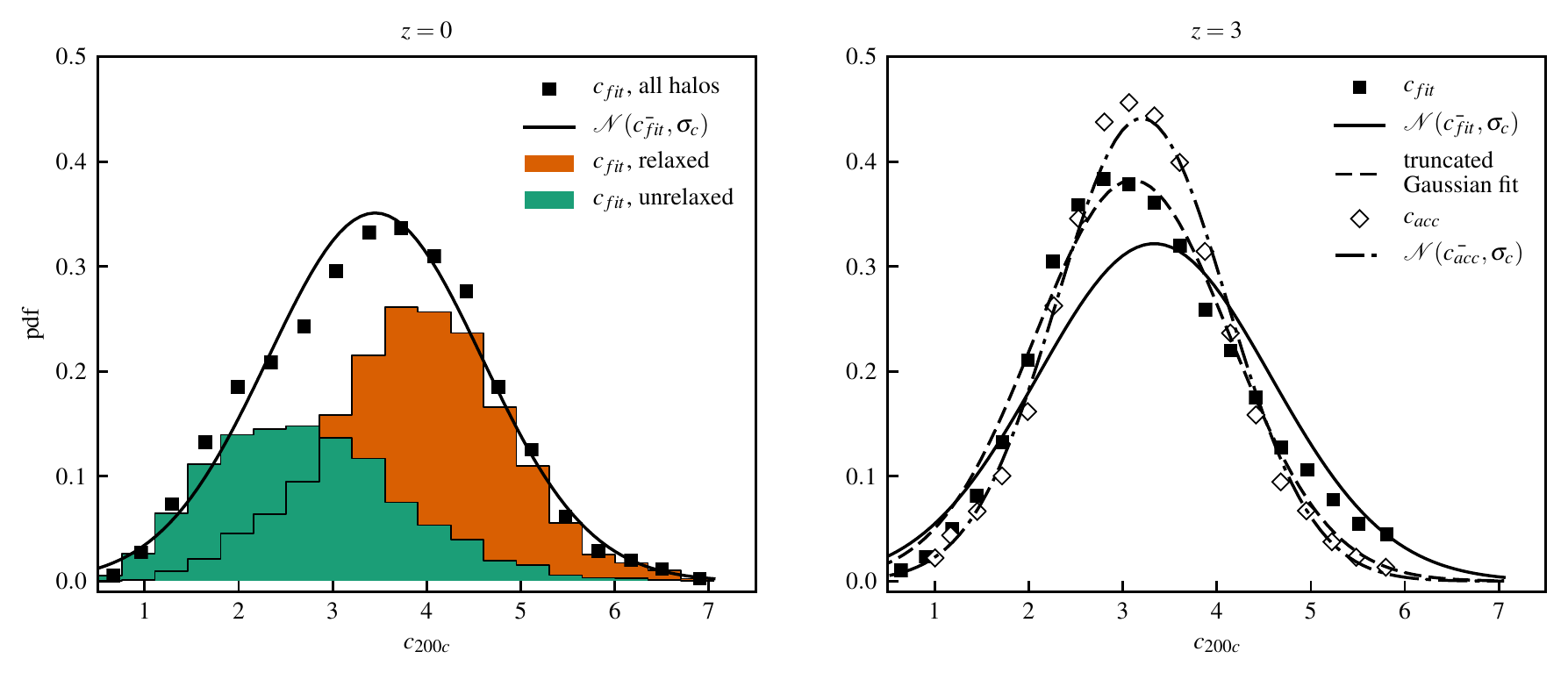}
    }
	\caption{Distributions of Outer Rim fit and accumulated mass concentrations. {\it Left panel:} fit concentrations are normally distributed in a high mass, low-redshift bin, $M_{200c} = 5.07 \times 10^{14} \: h^{-1}{ M}_\odot$ at $z=0$; the excess at low concentrations is due to unrelaxed halos (in this high-mass bin, 40\% of halos are unrelaxed). {\it Right panel:} high-redshift ($z=3$) bin centered at $M_{200c} = 1.08 \times 10^{13} \: h^{-1}{M}_\odot$, all halos (relaxed and unrelaxed). Solid and dot--dash lines show normal distributions with the sample mean and standard deviation, while dashed line is fit to the truncated $c_{\rm fit}$ distribution shown. In this bin, the distribution of fit concentrations is positively skewed, while the distribution of accumulated mass concentrations is not. }
\label{fig:conc_gaussian}
\end{figure*}

At all mass ranges and redshifts with sufficient halos in a bin, the distribution of accumulated mass concentrations is approximately normal, as shown in Figure~\ref{fig:conc_gaussian}. The accumulated mass scale radius can only fall between the innermost bin in the profile and $r_{200}$, limiting the range of $c_{\rm acc}$. However, fit concentrations can be arbitrarily high, so the distribution of $c_{\rm fit}$ is positively skewed, particularly at masses and redshifts where profiles are less well-described by the NFW form. See also Appendix~\ref{app:gauss_dstbn}.

\subsection{Stacked Halos}
\label{sec:stacked_halos}
Individual halo profiles can be noisy, but with 10 million halos of at least 2000 particles at $z=0$ in Q~Continuum and 20 million in Outer Rim, we can stack thousands of halos in a narrow mass bin (for example, $M \pm 1\%$, hereafter ``a 1\% stack'') to obtain smooth profiles. Relaxed halos are stacked by interpolating and summing the individual enclosed mass profiles. The differential mass profile is calculated from the mean enclosed mass profile. Note that we fit the 3D halo profile; a future paper will compare the concentrations that would be found for the same stacks using observational methods like weak lensing, which must fit the 2D projected mass profile.

Assuming the particles in our simulated profiles are drawn from a true NFW distribution, the stacked profile is a sum of NFW profiles with different concentrations, which, in principle, is not describable as an NFW profile. To visualize how the stacked profiles compare to an NFW profile, we calculate the effective power-law index of $\rho(r)$, that is, the slope of $\textrm{ln}\:\rho\left(\textrm{ln}\:r\right)$. For an NFW profile, 
\begin{equation}
n_\textrm{eff}^\textrm{NFW} = \frac{d \: \textrm{ln} \: \rho^\textrm{NFW}}{d \: \textrm{ln} \: (r/r_s)} = -\frac{1 + 3r/r_s}{1+r/r_s};
\end{equation}
the density is proportional to $r^{-1}$ at small $r$ and $r^{-3}$ at large $r$, crossing $n_\textrm{eff}^\textrm{NFW}(r_s) = -2$ at the scale radius. We calculate $n_\textrm{eff}(r)$ for stacked halos using a low-noise Lanczos differentiator with $N = 5$,
\begin{equation}
f'(x_i) \approx \frac{1}{10h}(-2y_{-2} - y_{-1} + y_1 + 2y_2);
\end{equation}
for the first two and last two bins,
\begin{eqnarray}
f'(x_0) \approx \frac{1}{6h}(-11y_0+18y_1-9y_2+2y_3), \\
f'(x_1) \approx \frac{1}{6h}(-2y_0-3y_1+6y_2-y_3).
\end{eqnarray}

\begin{figure*}[t]
	\centerline{
    \includegraphics{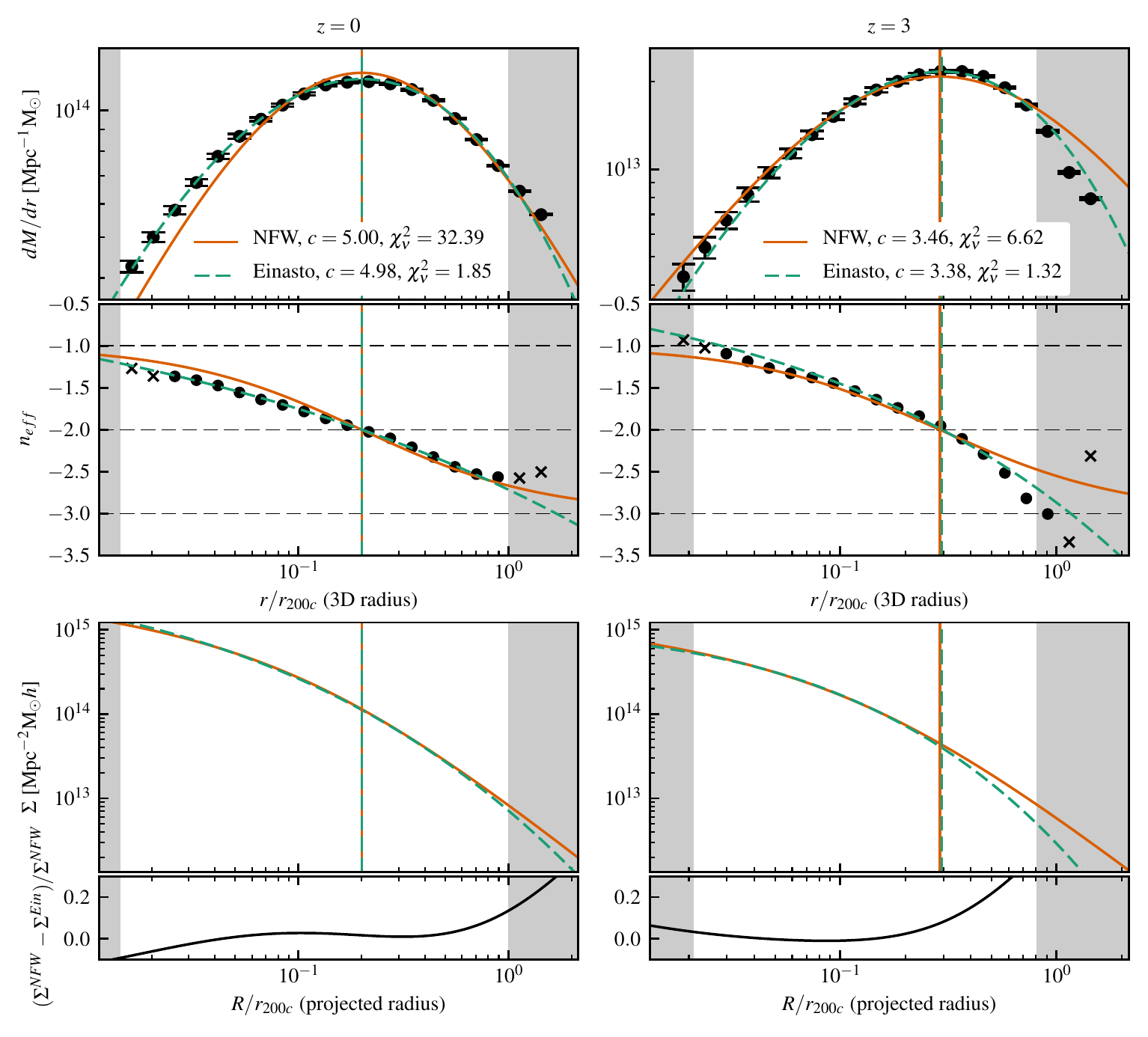}}
	\caption{\label{fig:stack1} Stacks of relaxed Q Continuum halos at $z=0$ (left panels, 116 halos, $M_{200c} = 6 \times 10^{13} \: h^{-1}{ M}_\odot \pm 0.2\%$) and $z=3$ (right, 81 halos, $M_{200c} = 9 \times 10^{12} \: h^{-1}{ M}_\odot \pm 1 \%$). Shaded regions are not included in the fit ($r > r_{200}$ or fewer than 100 particles enclosed); reduced $\chi_\nu^2$ values are calculated only on the points used to fit. {\it Top row}: NFW and Einasto profiles are fit to the $dM/dr$ profile as described in Section \ref{sec:conc}; vertical lines show the corresponding scale radii. At $z=0$, the NFW fit concentration is $c_\textrm{NFW} = 5.00$; Einasto fit concentration is $c_\textrm{Ein}=4.98$. At $z=3$,  $c_\textrm{NFW}=3.46$ and $c_\textrm{Ein}=3.38$. The Einasto fit captures the peak better than the NFW fit does at both redshifts; at $z=3$, it also improves on the high-$r$ behavior of the NFW profile. {\it Second row:} effective power-law index of the density profile. Slopes for the first and last two radial bins ($\times$ symbols) are less reliable than those with four neighboring points to include in the calculation. At high redshifts, our stacked profile is steeper than an NFW profile at high $r$. {\it Bottom rows:} surface mass density corresponding to the fit NFW and Einasto profiles and their relative difference; note that differences in the projected profiles are small, especially in the high-mass $z=0$ case.}
\end{figure*}

\begin{figure}
	\centerline{
    \includegraphics{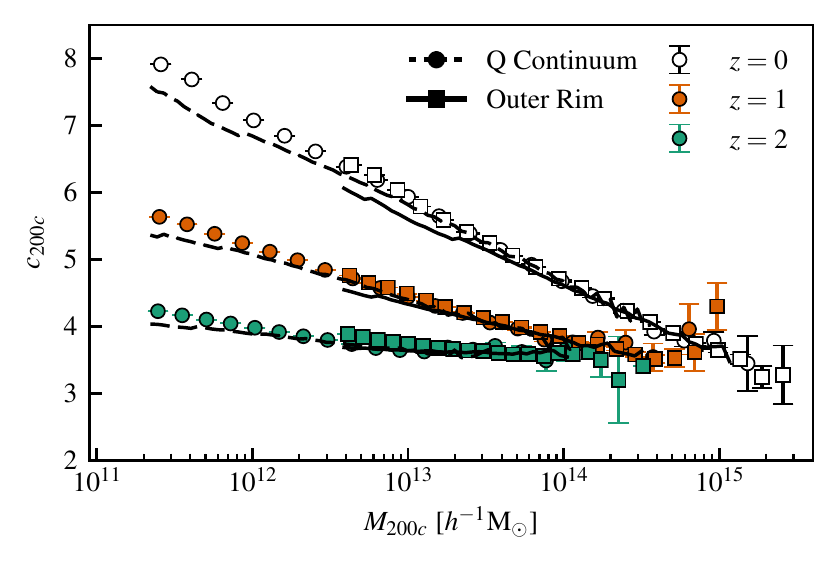}}
	\caption{\label{fig:stack2}Concentrations found using the profile fit method, $c_{\rm fit}$, for relaxed individual and stacked halos from both simulations (circles are for Q~Continuum, squares for Outer Rim). Points are means of individual fits; lines show concentrations found by fitting 5\% stacks at the same redshifts to an NFW profile.}
\end{figure}

Figure \ref{fig:stack1} shows examples of stacked profiles and their $n_{\rm eff}$ at $z=0$ and $z=3$. At high redshifts, the outer and inner profiles diverge from an NFW profile: by $r \sim r_{200c}$, density falls off more steeply than the NFW profile, while at small $r$, the density profile becomes shallower than the NFW profile (due to limitations imposed by force and mass resolution; however, results from our two simulations are still in good agreement, indicating little effect on the concentration itself). When we nevertheless calculate the NFW concentrations of the stacks, the three methods can find different concentrations, despite their agreement on the average concentration of individual halos. Additionally, the stacked fit concentrations differ from mean individual fit concentrations, as shown in Figure \ref{fig:stack2}, but by no more than 5\%,. The stacked concentrations are lower than individual means at low mass, but higher at high mass. These discrepancies reflect the fact that these stacked profiles deviate from the NFW form. 

\begin{figure}[htb]
	\centerline{
   	\includegraphics{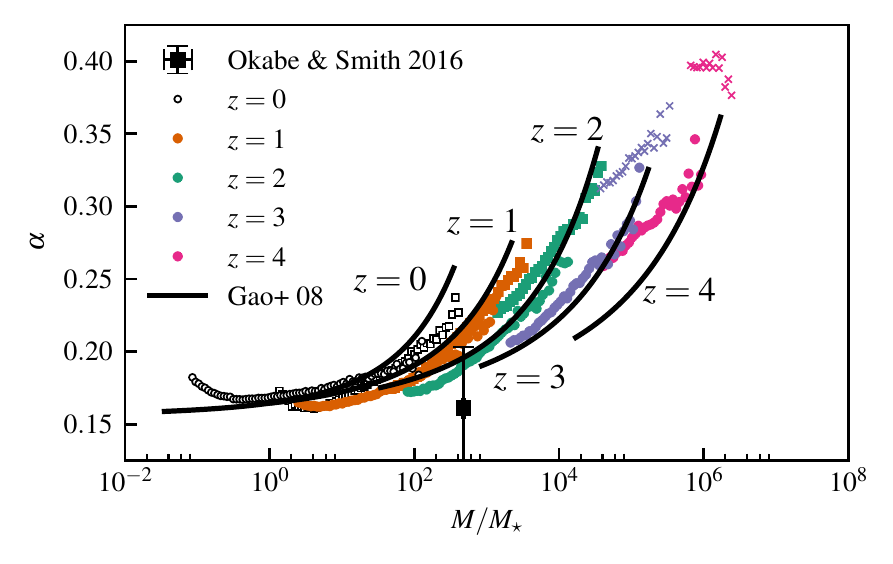}}
	\caption{\label{fig:alpha_mmstar} Relaxed halo Einasto $\alpha - M/M_\star$ relation for Q~Continuum (circles) and Outer Rim (squares). Einasto profiles are fit to halos stacked in 5\% mass bins, fixing mass to the SOD mass. High mass resolution is critical to measure the shape parameter $\alpha$; at high $z$, Outer Rim halos are insufficiently resolved ($\times$ symbols; see Appendix \ref{app:inner_profile} for discussion of high-redshift discrepancies between the two simulations). Black curves show the \cite{gao08} fit for $z = 0, 1, 2, 3, 4$. The single observational point is from a LoCuSS weak lensing measurement at $z=0.23$~\citep{okabe16}.}
\end{figure}

Previous studies have shown that profiles with a third parameter can better fit stacked profiles ~\citep{navarro04, merritt06, prada06, gao08}. After investigating several three-parameter profiles, we find the Einasto profile,
\begin{equation}
\textrm{ln}\left(\frac{\rho_E(r)}{\rho_{-2}}\right) = -\frac{2}{\alpha}\left\lbrack \left(\frac{r}{r_{-2}}\right)^\alpha -1 \right\rbrack,
\end{equation}
to be the best. We fit stacked halos to the Einasto form, allowing the concentration and the shape parameter $\alpha$ to vary freely. The mass $M_\Delta$ is the mean mass of halos in the stack, as determined by the SO algorithm. We find that the Einasto shape parameter increases with mass and redshift, as shown in Figure~\ref{fig:alpha_mmstar}. These results are in reasonable agreement with the model of \cite{gao08} fit to results at $0 \leq z \leq 3$.
The Einasto concentrations agree between the Outer Rim and Q Continuum runs, but the shape parameters differ at high redshift. See Appendix \ref{app:inner_profile} for further discussion of this discrepancy, and Appendix \ref{app:nu} for further comparison to other works.

Our results show that, at low mass, the stacked peak is to the left of the peak of the NFW profile fit to the stack, while at high masses at high redshift the opposite is true. At $z=0$, Einasto concentrations are greater than NFW concentrations at low mass; the Einasto and NFW $c$--$M$ relations cross around $M_{200c} = 10^{14} \: h^{-1}{M}_\odot$ (see Figure~\ref{fig:stack1} for an example stack whose peak location is well captured by the NFW fit), and at the highest masses, the Einasto concentration begins to fall below the NFW concentration. Luckily, at cluster scales at $z=0$, the NFW fit peak is transitioning from the left of the stacked peak to the right, and appears to be temporarily in agreement -- so NFW and Einasto profiles both fit the stacks very well, as found for the CLASH dataset by \cite{umetsu17}. These authors stack profiles in a wide mass bin; we do not find substantial differences in the quality of NFW vs. Einasto fits for wide mass bins, of up to 70\%, compared to narrower 1\% or 5\% bins.

\section{The Concentration--Mass Relation}
\label{sec:cm-rel}
The concentration measurements described above are now used to investigate the $c$--$M$ relation. As shown in Figures~\ref{fig:methods1} and \ref{fig:stack2}, the $c$--$M$ relation flattens with increasing redshift. At higher redshifts, halos of all masses have concentration $c \sim 3-4$ and the $c$--$M$ relation falls no further; \cite{zhao03} found a similar floor. We do not present results for redshifts higher than $z=4$ (see discussion in the appendix), but concentrations measured up to redshifts as high as $z=10$ are consistent with this floor.

In order to represent the data in a $z$-independent form, we scale the mass by the redshift-dependent nonlinear (or ``collapse'') mass scale $M_\star = 4\pi/3 \: \rho_c(z)\: \omega_m(z) \: R_\star^{3}$, which solves
\begin{equation}
\sigma(R_\star,\:z) = \delta_c,
\end{equation}
where $\delta_c = 1.686$ is the linear critical density for collapse and $\sigma(M,z)$, the amplitude of mass fluctuations,  is the power spectrum smoothed with a top-hat filter. The power spectrum $P(k,\:z) = d^2(a) \: P(k, \:z=0)$ is calculated from the growth factor
\begin{equation}
d(a) = \frac{D^+(a)}{D^+(a=1)},
\end{equation}
\begin{equation}
D^+(a) = \frac{5 \Omega_m}{2} \frac{H(a)}{H_0} \int_0^a \frac{da'}{[a' H(a') /H_0]^3}
\end{equation}
and $P(k, z=0)$ is as given by CAMB \citep{lewis00}. The mean square perturbation is
\begin{equation}
\label{eqn:sigma}
\sigma^2\left(R,z\right)=\frac{1}{2\pi^2}\int k^2 \: dk \: W^2\left(k,R\right)\:P\left(k,z\right),
\end{equation}
where we choose the window function $W^2(k, R)$ to be the Fourier transform of a spherical top-hat filter of radius $R$,
\begin{equation}
W(k, R) = \frac{3}{(kR)^3}[\textrm{sin}(kR)-kR \: \textrm{cos}(kR)].
\end{equation}

\begin{figure*}[htb]
	\centerline{
   	\includegraphics{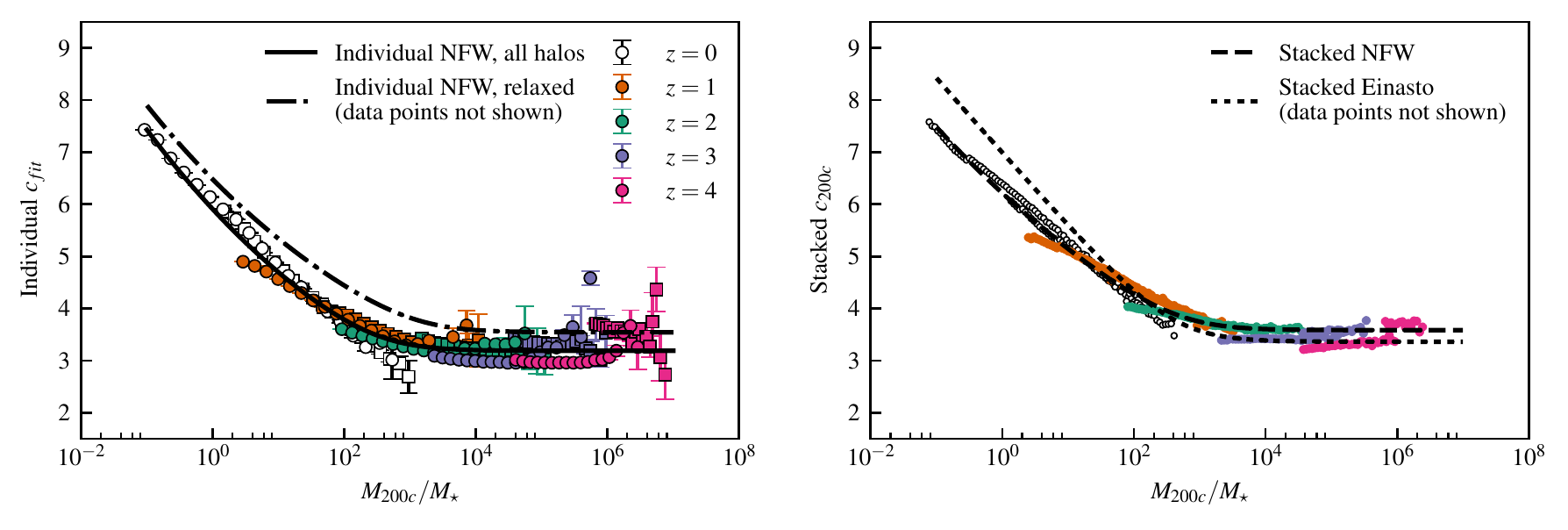}
    }
	\caption{Individual mean $c_\textrm{fit}$ for all halos (left) and stacked $c_\textrm{fit}^\textrm{NFW}$ for relaxed halos in 5\% stacks (right) as a function of $M/M_*$ across eight orders of magnitude for Q Continuum (circles) and Outer Rim (squares). Red solid (left) and dashed (right) curves are fit to all points shown, as well as $z=0.502$ and $z=1.494$. Dotted--dashed red curve is fit to individual mean concentrations of relaxed halos only (left panel, data points suppressed to avoid clutter); dotted is fit to Einasto concentrations of 5\% stacks (right panel, data points suppressed). }
\label{fig:cm_mmstar} 
\end{figure*}

The nonlinear mass $M_\star$ depends weakly on cosmology and falls steeply with redshift: for our cosmology, $\textrm{log}(M_\star / h^{-1} { M}_\odot) = 12.5$ at $z=0$, $11$ at $z=1$, and $9.5$ at $z=2$. Already at $z=3$,  $\textrm{log}(M_\star / h^{-1} { M}_\odot) = 8$, and our least-massive halos exceed the nonlinear mass scale by three orders of magnitude at this redshift. Combining the results of the two simulations at redshifts up to $z=4$, we probe eight decades in $M/M_\star$, as shown in Figure~\ref{fig:cm_mmstar}. 

When represented in terms of $M/M_\star$, the individual $c$--$M$ relations at different redshifts fall relatively tightly onto a single relation: the concentration behaves as a power law with $M/M_\star$ as long as $M/M_\star \lesssim 10^3$; at masses farther above the nonlinear mass scale, the concentration asymptotes to a constant value $\sim 3$. This behavior is shown in Figure~\ref{fig:cm_mmstar} for both individual and stacked halos, and is consistent with analytic models which find concentration to be independent of mass above a threshold mass. \cite{dalal10} find concentration to approach a constant $c \sim 4$ for high-mass halos, those for which $\sigma(M) \ll \delta_c$ (implying $M \gg M_\star$), where Gaussian statistics are a good description of the slopes of the corresponding peaks in the initial Gaussian random field. \cite{okoli16}, similarly, predict a constant $c$--$M$ relation, $c \sim 2.5$, in the high-mass regime where an assumption of spherical collapse is valid. See further discussion in Appendix~\ref{app:cm_others}.

We fit to a simple functional form that captures this behavior:
\begin{equation}
\label{eqn:cmmstar_form}
c_{200c} = A\left\lbrack\left(\frac{M_{200c}/M_\star}{b} \right)^{m} \left( 1+\frac{M_{200c}/M_\star}{b}\right)^{-m} - 1\right\rbrack + c_0,
\end{equation}
transitioning at $M=M_T\equiv b M_\star$ from a power law to a constant $c=c_0$. This plateau is found at a concentration between 3 and 4, depending on the halos included (all halos vs. relaxed halos only, for example) and the type of fit. Table \ref{tab:fit_cmmstar} gives the fit parameters. We note that the fit should not be na\"{i}vely extrapolated to masses smaller than those considered here.

This form is not fully universal in the sense of being approximately cosmology-independent, as can be shown by checking against the $w$CDM results of \cite{bhattacharya13} and \cite{kwan13}. We are currently investigating the detailed cosmology dependence of the $c$--$M$ relation using the Mira--Titan Universe suite of simulations \citep{heitmann16, lawrence17}, which cover $\sim$100 cosmological models allowing for a $(w_0,w_a)$ parameterization of dark energy as well as the effect of massive neutrinos. These results will be presented elsewhere.

The peak-height parameter $\nu$ is frequently used to find a redshift-independent $c$--$\nu$ relation, with similar motivations (see, e.g. \citealt{bhattacharya13, dutton14, ludlow14, diemer15}).  We find our results to be more universal with redshift as a function of $M/M_\star$ than as a function of $\nu$ (see Appendix \ref{app:nu} for further details). Moreover, we find $z$-independence to hold for other overdensities, although it does deteriorate for smaller choices of $\Delta$ (including $\Delta_{\rm vir}$).

\begin{table}
\centering
\resizebox{1.0\columnwidth}{!}{
\begin{minipage}{1.0\columnwidth}
\centering
\caption{{\normalfont $c$--$M/M_\star$ fit parameters, $0 \leq z \leq 4$}}
\label{tab:fit_cmmstar} 
\begin{tabular}{c c  c  c  c } 
\tableline \tableline
 Type of Fit & $m$ & $A$ & $M_T/M_*$ & $c_0$ \\ \tableline
Individual, all & $-0.10$ & $3.44$ & $430.49$ & $3.19$ \\  
Individual, relaxed & $-0.09$ & $2.88$ & $1644.53$ & $3.54$\\  
Stack, NFW & $-0.07$ & $4.61$  & $638.65$ & $3.59$ \\
Stack, Einasto & $-0.01$ & $63.2$  & $431.48$ & $3.36$\\ \tableline
\multicolumn{5}{c}{{\bf Notes.} Use Equation~\ref{eqn:cmmstar_form} with variance $\sigma_c = c_{200c}/3$.}
\end{tabular}
\end{minipage}}
\end{table}

\begin{figure}[htb]
	\centerline{
   	\includegraphics{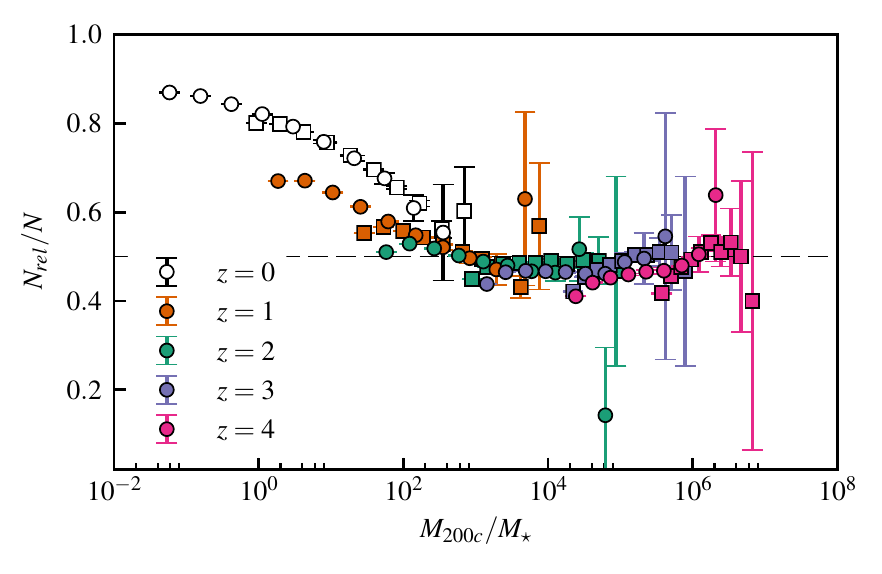}
    }
	\caption{The fraction of Q~Continuum (circles) and Outer Rim (squares) halos considered relaxed (distance between most bound particle and FOF center of mass at most $0.07R_{200}$) falls with mass until $M\sim M_T$, but is approximately constant at $0.5$ for $M>M_T$.}
\label{fig:relaxed_mmstar} 
\end{figure}

Aside from the concentration, it is also important to investigate the fraction of relaxed halos as a function of $M/M_\star$. Because halos with $M\gg M_\star$ are likely to be in the ``halo formation'' phase, the unrelaxed fraction should increase with mass. However, one would anticipate that the existence of a finite asymptotic value of the concentration, $c_0$, suggests that the relaxed fraction also reaches a limiting value. We find that, like the concentration, the relaxed fraction is also independent of mass for $M_{200c} > 10^{11}{ M}_\odot$ at redshifts higher than $z \sim 2$, or alternatively, for $M/M_\star \gtrsim 10^3$, as shown in Figure~\ref{fig:relaxed_mmstar}. At low redshift, up to 80\% of low-mass halos are relaxed, falling to 50\% at high masses, which is still a substantial fraction; at high redshifts, 50\% of halos (that pass our minimum mass threshold) are relaxed at all masses, even for $M\gg M_\star$. 

\subsection{Comparison with Observations}
\label{subsec:obs}

We compare our $c$--$M$ relation at $z=0$ to individual and stacked halo  concentrations measured from individual X-ray \citep{vikhlinin06}, individual weak and strong lensing \citep{newman13, merten15}, and stacked lensing \citep{mandelbaum06, mandelbaum08, johnston07, okabe10, okabe13, umetsu11, umetsu14, umetsu16, oguri12, brimioulle13, sereno13, foex14, niikura15,  okabe16, cibirka17} observations. Concentrations and masses are reported in a variety of mass definitions ($M_{500c}$,  $M_{200c}$, $M_{\rm vir}$); when $M_{200c}$ is not provided, we assume an NFW profile and convert to $M_{200c}$, $c_{200c}$ according to \cite{hu03}.

All observations considered here were carried out at redshifts between $z=0$ and $z=1$; in the case of stacked analyses, we assign a mean redshift to the sample. We fit the $c$--$M$ relation at $z=0, 1$ to 
\begin{equation}
\label{eqn:cm_powerlaw}
c_{200c} = A \left(1+z\right)^d M^m;
\end{equation}
the fit parameters are shown in Table \ref{tab:fit_cm}. Note that this simple fit is only valid for redshifts $0 \leq z \leq 1$, where a power-law dependence on mass and redshift is a reasonable description of our results. Individual (Figure \ref{fig:obs_indiv}) and stacked (Figure \ref{fig:obs_ensemble}) measurements are scaled to $z=0$ according to the fit redshift dependence and compared to the corresponding mean $c_{\rm fit}$ of individual relaxed halos or stacked $c_{\rm fit}$. Note that our Einasto and NFW fit concentrations are identical around $M_{200c} \sim 10^{14} \: h^{-1}{ M}_\odot$, and are only beginning to diverge at cluster scales, as mentioned in Section~\ref{sec:stacked_halos} above. This is the mass range where, as shown in Figure~\ref{fig:stack1}, Einasto and NFW profiles both provide very good fits and are difficult to distinguish. The CLASH measurement of \cite{umetsu17}, where both Einasto and NFW forms are able to fit the projected mass profile, also falls in this range.

The current agreements with observations are very good for both individual and stacked halo measurements, although there are a few outliers. Next-generation CMB, optical, and X-ray surveys will greatly increase the number of well-measured group and cluster-scale halos. This will lead to much better control on mass measurement for stacked observations (currently at $4\%$, as claimed by LoCuSS), as well as on the profile measurements (\citealt{simet17}).

\begin{table}
\centering
\resizebox{1.0\columnwidth}{!}{
\begin{minipage}{1.0\columnwidth}
\centering
\caption{{\normalfont Power-law $c$--$M$ fit parameters, $0 \leq z \leq 1$ }} 
\label{tab:fit_cm}
\begin{tabular}{c c  c  c } 
\tableline \tableline
 Type of fit & $A$ & $d$ & $m$ \\ \tableline
individual, all & $75.4$ & $-0.422$ & $-0.089$ \\   
individual, relaxed & $68.4$ & $-0.347$ & $-0.083$ \\  
stack, NFW & $57.6$ & $-0.376$  & $-0.078$ \\
stack, Einasto & $122$ & $-0.446$  & $-0.101$ \\ \tableline
\multicolumn{4}{c}{{\bf Notes.}  Use Equation~\ref{eqn:cm_powerlaw} with variance $\sigma_c = c_{200c}/3$.}
\end{tabular}
\end{minipage}}
\end{table}

\section{Discussion}
\label{sec:disc}
The high-volume, high-resolution simulations Q Continuum and Outer Rim simultaneously provide superior statistics (20 million and 10 million halos of at least 2000 particles at $z=0$, respectively) and halo resolution, allowing us to use multiple methods of concentration measurement to study the $c$--$M$ relation across eight decades in $M/M_\star$, the distribution of concentration within mass bins, and smooth stacked profiles of at least hundreds of halos in narrow (5\% and smaller) mass bins. 

Our three methods of concentration measurement (fit, accumulated mass, and peak finding) agree best on massive halos at low redshift; at higher $z$, mean measurements differ by 10--20\%. This is consistent with the general expectation that concentration measurements can have systematic differences depending on how the measurements are carried out (\citealt{bhattacharya13,meneghetti13,dutton14}; see Appendix \ref{app:cm_others} for further discussion). These caveats aside, our results are in excellent agreement with observations as well as with most recent simulations.

The $c$--$M$ relation is not a precise, narrow correlation between the halo concentration and halo mass. There is a substantial amount of intrinsic variability, which appears, remarkably, to be cosmology-independent; it is specified by $\sigma/c\sim 1/3$, where $\sigma$ is the standard deviation around the mean concentration at a given halo mass (\citealt{dolag04,bhattacharya13}). Our results are consistent with a Gaussian distribution of concentration within mass bins (\citealt{lukic09, reed11, bhattacharya13}). Small deviations at lower concentrations can be explained by an unrelaxed halo population, and at higher concentrations due to the existence of a small high-concentration tail.

Because of the excellent statistics made possible by our simulations, we can study the stacked profiles of halos in narrow mass bins. We find that the Einasto profile is an excellent fit to the data, and that concentrations measured from our stacked profiles agree with the means of the concentrations of those halos measured individually. 

Our well-characterized results for the $c$--$M$ relation across a wide range of redshifts motivate searching for a simple description of the data in a redshift-independent form, much as in the case of the halo mass function. Although there is no basic theory for the $c$--$M$ relation and its evolution with redshift, there exist a number of models and fits to numerical data. These include some simple analytical ideas, power-law models, and significantly more complex fits (see Appendix \ref{app:other_models}). We find that scaling the halo mass by the nonlinear mass scale $M_\star$ allows us to describe results from different redshifts by a single relation with a very simple form. This $c$--$M$ relation has a power-law form at lower masses (over the mass range investigated) and transitions to a constant value at masses above a threshold mass $M_T\sim 500-1000M_\star$. At redshifts higher than $z=3$, all halos above the threshold of 2000 simulation particles exceed $M_T$ and have concentration $c\sim 3$. 
At low redshifts (such as $z=0$, where $M_T\sim 5-10\times 10^{14.5}\:{ M}_\odot$), only the most massive (cluster-scale) halos approach $M_T$. These halos, too, have concentrations $c \sim 3$. 

Our results are in excellent agreement with current observational datasets. Future measurements will have significantly enhanced statistics for both individual and stacked halo profiles. This will bring the observational errors closer to the current error estimates from the simulations and allow for any differences to become more apparent. Increasing our understanding of the halo profiles will aid in controlling systematic errors in a number of cosmological analyses. In an upcoming paper, we will present halo mass accretion histories and discuss their implications for the shape and redshift-dependence of the $c$--$M$ relation. We are also currently investigating halo profiles in the Mira--Titan suite of simulations \citep{heitmann16}, to be able to predict deviations from $\Lambda$CDM.


\acknowledgements
\section*{Acknowledgements}
The authors thank members of the Cosmological Physics and Advanced Computing Group at Argonne for many useful discussions. S.H. acknowledges helpful exchanges with Rachel Mandelbaum. 

Argonne National Laboratory's work was supported under the U.S. 
Department of Energy contract DE-AC02-06CH11357. The authors acknowledge support from the ASCR/HEP SciDAC program. This research used
resources of the ALCF, which is supported by the DOE Office of Science under contract
DE-AC02-06CH11357 and resources of the OLCF, which is supported by
DOE/SC under contract DE-AC05-00OR22725. Some of the results presented
here were made possible by awards of computer time provided by the ASCR
Leadership Computing Challenge (ALCC), the ASCR Innovative and Novel Computational Impact on Theory and Experiment (INCITE) programs at ALCF and OLCF, and an Early Science Project at the ALCF. 

The CosmoSim database used in this paper is a service provided by the Leibniz-Institute for Astrophysics Potsdam (AIP).
The MultiDark database was developed in cooperation with the Spanish MultiDark Consolider Project CSD2009-00064.

The work of H.C. was supported by the National Science Foundation Graduate Research Fellowship Program under Grants No. DGE-1144082 and DGE-1746045.


\appendix

In this appendix, we compare our results with those from other simulations commenting on agreements and differences. We also investigate various simulation errors that could affect halo concentrations, demonstrating that our results are robust to these potential error sources. We note that the Outer Rim and Q~Continuum simulations have mass resolutions that are different by a factor of 20 and were run using different short-range force algorithms ($P^3M$ versus tree). Achieving very consistent results, given these differences, is itself a good test of the overall methodology. 

\section{Other Simulations}
Our results are in reasonably good agreement with those of recent simulations; differences often arise due to choices made in the operational definition of the concentration.

\subsection{Profile Shape}
\begin{figure}[htb]
	\centering
   	\includegraphics{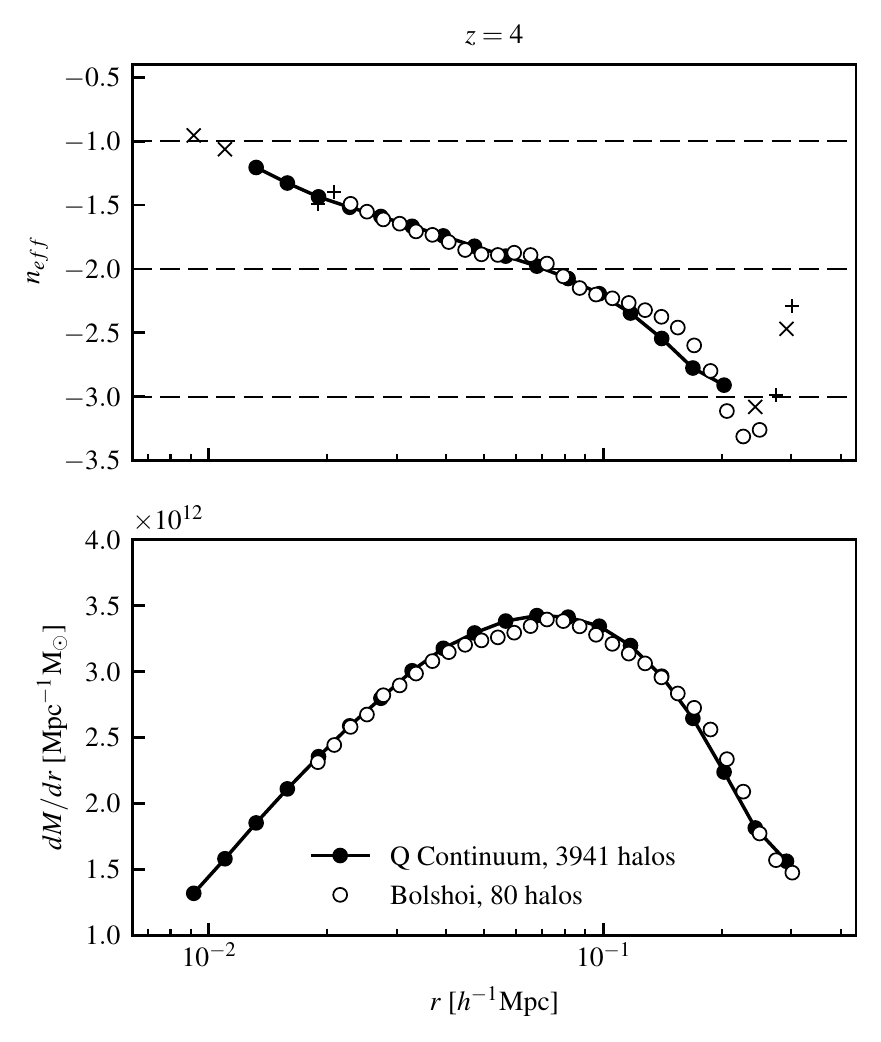}
	\caption{Stacked Q Continuum and Bolshoi profiles in a narrow mass bin at $z=4$, $M_{200c} = 6 \times 10^{11} \: h^{-1}{ M}_\odot \pm 1\%$, relaxed halos only. Slopes for the first and last two radial bins (shown as $+$ symbols for Bolshoi, $\times$ for Q~Continuum) are less trustworthy than those calculated from the full five points.}
\label{fig:qc_bolshoi}
\end{figure}

To illustrate the closeness of obtained profiles across simulations, we compare our results with \cite{klypin11} and \cite{prada12} who  calculate the concentrations of halos from the MultiDark and Bolshoi project suite of simulations. These halo profiles are publicly available in the CosmoSim database\footnote{https://www.cosmosim.org/cms/simulations}. The cosmologies of the Bolshoi and MultiDark simulations differ from ours, but the mass resolutions of Bolshoi ($1.35 \times 10^8 \: h^{-1} { M}_\odot$) and Q Continuum ($1.05 \times 10^8 \: h^{-1}{ M}_\odot$) are similar. A stacked profile of high-redshift Bolshoi halos overlaps with a Q Continuum profile at the same mass and redshift (Figure \ref{fig:qc_bolshoi}); the shape of the halo profiles is essentially the same.

\subsection{The $c$--$M$ Relation in Other Simulations}
\label{app:cm_others}
\begin{figure*}
	\centerline{
    \includegraphics{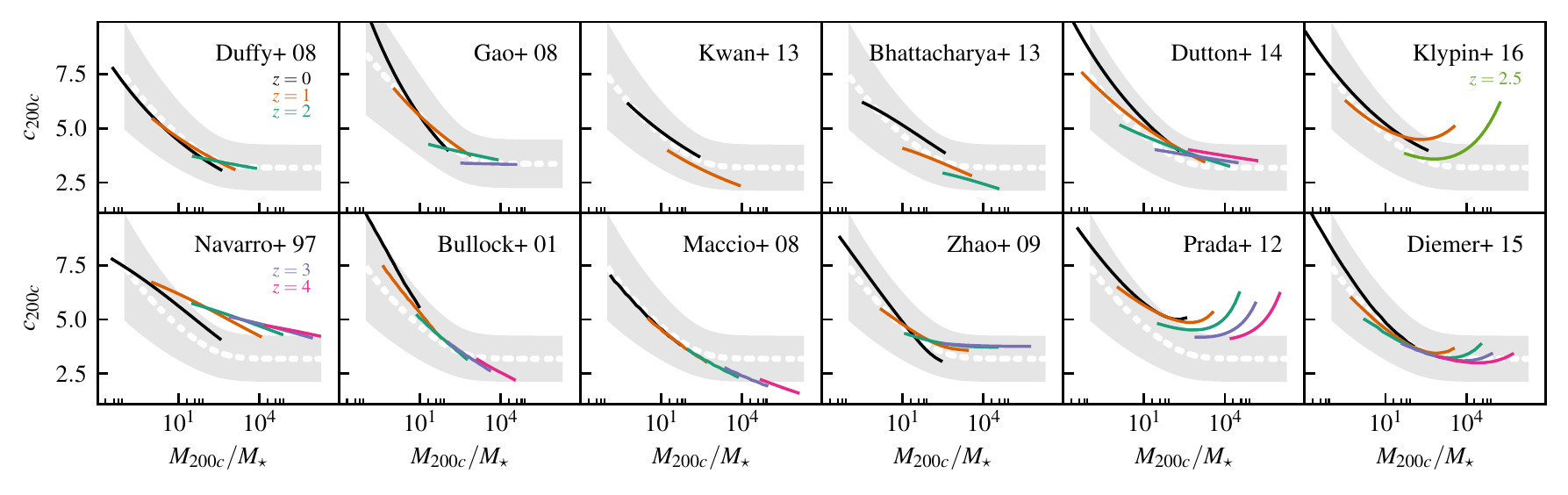}
    }
	\caption{\label{fig:cm_others_separate} Other fits and models generally fall within one standard deviation (shaded region, $\sigma_c = c/3$) from our fit to concentrations at $0 \leq z \leq 4$ (white, dashed); all but \cite{gao08} are results for all (relaxed and unrelaxed) individual halos. Note that each panel shows results from multiple redshifts, to illustrate redshift-independent behavior when mass is scaled by $M_\star$. \textit{Top row:} fits. \cite{duffy08} (WMAP-5 cosmology), \cite{gao08} ($\sigma_8 = 0.9$, Einasto concentrations of relaxed halos only), and \cite{dutton14} (Planck cosmology) results are power-law fits to cosmologies that differ from ours; $M_\star$ is calculated accordingly, but note that $M_\star$ scaling does not provide full cosmology independence. The other three fits are shown for our WMAP-7 cosmology. \textit{Bottom row:} models. The predictions of \cite{nfw2} (with the free parameter $f=0.01$), \cite{zhao09}, \cite{prada12}, and \cite{diemer15} are calculated for our cosmology. The cosmologies of \cite{bullock99} ($\sigma_8 = 1$; $F=0.01$, $K=4.0$) and \cite{maccio08} (WMAP-5; $F=0.01$; $K=3.6$) differ from ours; $M_\star$ is calculated accordingly. See text for further description of the models.}
\end{figure*}

Many different methods have been used to measure concentrations from a large set of simulations. The results fall in one of three categories: concentration is power-law at all masses and redshifts; concentration flattens at high masses and redshifts; or concentration increases at high mass and redshift. Our results are most consistent with the second category; we do not find evidence that concentration increases with mass at high masses and redshifts, but instead find that it approaches a constant $c \sim 3$. Selected fits and analytical models of all three types are compared to our results in Figure \ref{fig:cm_others_separate}. Masses are scaled by $M_\star$ and multiple redshifts are shown to check for redshift-independent behavior in others' results; note, however, that the $c$--$M/M_\star$ relation is not fully cosmology-independent -- concentrations are higher in high-$\sigma_8$ cosmologies. 

The first papers to note a deviation from power-law behavior at high mass and redshift described a flattening of the $c$--$M$ relation. \cite{zhao03} describe a concentration floor at $c \sim 3$ and a mass-independent $c$--$M$ relation at high redshift; their $z=4$ plot may show a slight upward slope. \cite{neto07} provide results only for $z=0$, but comment on a tendency toward constant concentration at high mass. \cite{gao08} note a flattening at high mass and redshift. The model of \cite{zhao09}, which predicts concentration from the time when a halo reaches a fraction of its final mass, predicts a flat regime at high masses and redshifts; again, the results at $z=4$ may slope slightly upward, but are fit to a model that predicts mass-independent concentration. Two results at high redshift also find very shallow $c$--$M$ relations: \cite{hellwing16} provide results for relaxed halos only, but find that the $c$--$M$ relation flattens at redshifts up to $z=9$. \cite{angel16} find a shallow $c$--$M$ relation for relaxed halos at $z>5$; the $c$--$M$ relation for the full halo sample is shown to be consistent with the upturn of \cite{diemer15}, but also appears to be consistent with a flat relation.

Using a new method to measure concentration, based on maximum circular velocity $V_{\rm circ}$, \cite{klypin11} note an upturn in the $c$--$M$ relation. The upturn is stronger for relaxed halos (in addition to our criterion that the distance between halo centers fall within $0.07R_{200}$, their halos are only relaxed if spin parameter $\lambda < 0.1$). The reasons for the upturn are unknown, but the statistics of the highest density peaks are suggested as a possible explanation. Using a modified version of the $V_{max}$ method and including only bound particles in halo profiles, \cite{prada12} also find an upturn. The upturn is also seen in relaxed halos selected using the spin parameter and center-offset criteria as well as an additional requirement that the virial ratio $2K/|U| < 0.5$. Elaborating on the statistics of high-density peaks, \cite{prada12} show that infall velocity is greatest for the most massive halos. \cite{klypin16} returns to this point, suggesting that greater infall velocity will produce more concentrated halos, and notes a relationship between true concentration and the Einasto shape parameter $\alpha$. \cite{munozcuartas11} use a profile fitting method rather than $V_{max}$; power-law fits are used to describe their results, but a flattening of the $z=2$ $c$--$M$ relation at high mass is noted to be consistent with \cite{klypin11}. \cite{correa15c} fit density profiles, and find an upturn or flattening when all halos are included in the $c$--$M$ relation, but not when the sample is restricted to relaxed halos. \cite{diemer15} also use a density profile fitting method, find an upturn, and propose a model that predicts a positive slope for the high-$\nu$ $c$--$\nu$ relation.

The methods of the above papers differ in several respects. Other papers discuss the effects of these differences in methodology on the $c$--$M$ relation. \cite{ludlow12} find an upturn using the $V_{max}$ method, but the upturn is not seen when the halo sample is restricted to relaxed halos selected using two criteria (virial ratio and fraction of mass in substructure) in addition to our restriction on the distance between halo centers (the $c$--$M$ relation for our unrelaxed halo sample does not differ from that of relaxed halos in shape, only in amplitude). The additional criteria are relevant, as halos are seen to pass through phases of high concentration immediately following major mergers: when the infalling progenitor passes near the center of the halo, the halo is highly concentrated and is classified as unrelaxed according to the virial ratio test. \cite{ludlow14}, \cite{angel16}, and \cite{hellwing16} also see no upturn in the relaxed-halo $c$--$M$ relation. \cite{meneghetti13} are able to bring the results of \cite{prada12} into agreement with those of \cite{duffy08} by binning halos in $M_{200}$ as opposed to $V_{max}$, and by measuring concentration through profile fitting as opposed to the $V_{max}$ method. \cite{dutton14} do find that the $c$--$M$ relation slopes upward at $z=3$ when the $V_{max}$ method is used, but is flat when concentrations are determined by a profile fit method. The $c$--$M$ relations are fit to power laws, but do show either a slight upturn or plateau at high masses and redshifts. \cite{rodriguezpuebla16}, however, do find an upturn in concentrations measured by profile fitting as well as using the $V_{max}$ method; \cite{klypin16} find an upturn both when binning in $V_{max}$ and in virial mass.

\subsection{Models of the $c$--$M$ Relation}
\label{app:other_models}
Several different fits and models, with varying levels of physical motivation, have been used to describe the $c$--$M$ relation. The simplest description is a power-law fit at each redshift to concentration as a function of mass (top row of Figure \ref{fig:cm_others_separate}: \citealt{duffy08, gao08, kwan13, dutton14}).  In the case of the upturn, a third parameter is needed to fit $c(M)$, as in \cite{klypin16}. The peak-height parameter $\nu$ has been used as the independent parameter to find a redshift-independent $c$--$\nu$ relation; the shape may simply be fit, as in e.g. \cite{bhattacharya13}, or described by a model (bottom row of Figure~\ref{fig:cm_others_separate}).

One set of models assumes that the density of the central regions of a halo is determined by the background density at the time the halo forms. The model of \cite{nfw2} predicts the concentration of a halo of mass $M$ from the collapse redshift $z_{\rm coll}$, the redshift when the total mass of progenitors of mass $fM$ ($f<1$) or greater first reaches $0.5M$, calculated using the extended Press--Schechter formalism. Concentration is found by assuming that the mean density at $z_{\rm coll}$ determines the density of the final halo within $R_s$. The model is found to describe simulated $c$--$M$ relations for small values of $f \sim 0.01$. \cite{bullock99} revise the model of \cite{nfw2}; the collapse redshift is found as the time when the collapse mass, $M_\star$, first reaches a fraction of the final halo mass, $FM$. In addition to the free parameter, $F$, a second parameter, $K$, relates a density of the halo to the background density at $z_{\rm coll}$. \cite{maccio08} modify the \cite{bullock99} model, allowing the halo overdensity with respect to the mean background density to change with redshift. 

Using mass accretion histories, \cite{zhao09} find that concentration of a halo of mass $M$ can be calculated from the age of the universe, $t_{0.04}$, when its main progenitor first reached a mass of $0.04M$. This time is calculated from a model of mass accretion history. The fit to concentration as a function of $t$ and $t_{0.04}$ predicts a constant concentration of $c=4$ when $t/t_{0.04} < 3.75$. Combined with the mass accretion histories, this implies a flat $c$--$M$ relation at high redshift, where halos rapidly accrete mass; by $z \sim 3$, a halo of mass $10^{12} h^{-1} M_\odot$ is approaching the stage of its evolution where concentration begins to grow. This model is also used by \cite{vandenbosch14}, with modified parameters, to describe $z=0$ concentrations found using the maximum circular velocity method of \cite{klypin11}. \cite{wechsler02} and \cite{zhao03} also provide concentrations based on mass accretion history fits. \cite{correa15c} combine mass accretion histories, extended Press--Schechter theory, and a fit relationship between concentration and formation time, defined as the time the mass of a halo's main progenitor reaches the mass within the scale radius of the descendant halo. This model predicts concentration that falls with mass for relaxed halos at all masses and redshifts.

Two other models use an additional cosmology-dependent variable to fit the $c$--$\nu$ relation. \cite{prada12} describe concentration as a function of $\sigma$ and a time variable $x$, motivated by its appearance in the growth factor $D$. \cite{diemer15} develop a model in which concentration is a function of both $\nu$ and the local slope of the power spectrum, $n$. 

Two papers that predict halo concentrations from the initial Gaussian random field are consistent with our results. \cite{dalal10} predict halo profiles from the shapes of initial peaks in the Gaussian random field, then fit those profiles to an NFW form to find the concentration. The shape of the $c$--$M$ relation is determined by the outer slopes of these peaks. For high-mass halos, where slope is shallow and consistent with a prediction from Gaussian statistics, concentration $c \sim 4$ is independent of halo mass. The outer slopes of simulated peaks corresponding to low-mass halos differ from the Gaussian statistics predictions and produce halo profiles with higher concentrations. The transition between these two regimes is redshift-dependent, but falls between $M_\star$ and $1000M_\star$ for $0 \leq z \leq 4$. \cite{okoli16} calculate the initial energy of a region before collapse, then assume energy is conserved and use the Jeans equation to find concentration from the final energy. In the case of spherical collapse (valid for the largest halos), the resulting concentration $c \sim 2.5$ is mass-independent. For smaller halos, ellipsoidal collapse predicts concentrations that fall with mass. 

\subsection{Gaussian Distribution of Concentrations}
\label{app:gauss_dstbn}
\begin{figure}[htb]
	\centerline{
    \includegraphics{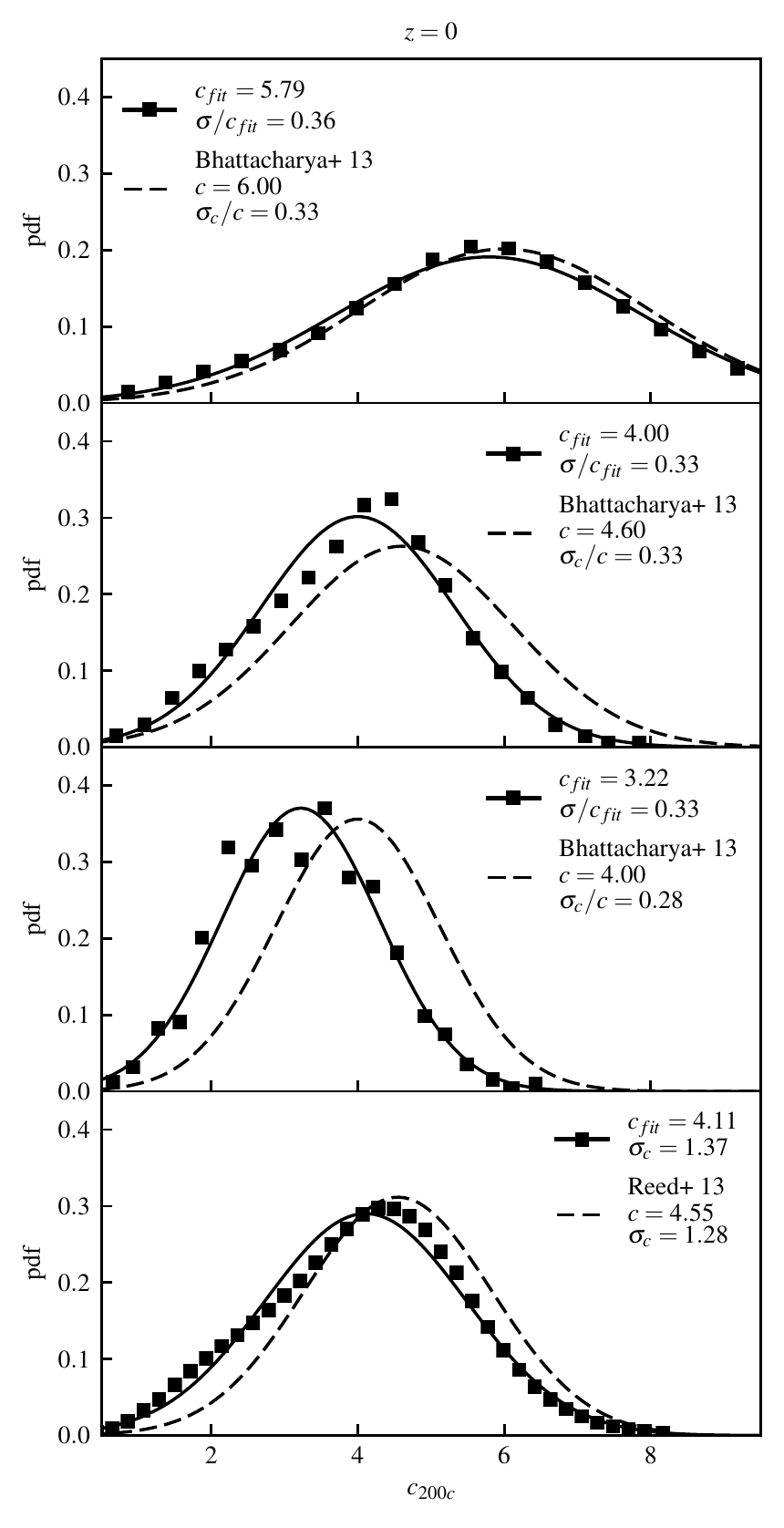}
    }
	\caption{\label{fig:conc_gaussian_reed} {\it Top three panels:} distributions of Outer Rim fit and accumulated mass concentrations for three mass bins at $z=0$: $M_{200c} = 5 \times 10^{12} \: h^{-1}{M}_\odot \pm 1 \%$ (top), $M_{200c} = 1.5 \times 10^{14} \: h^{-1}{M}_\odot \pm 1 \%$ (middle), and $M_{200c} = 8 \times 10^{14} \: h^{-1}{M}_\odot \pm 5 \%$ (bottom). All halos, relaxed and unrelaxed, are included. The results in these three panels are  analogous to those presented in Figure~A13 of \cite{bhattacharya13} (black dashed curve). {\it Bottom panel:} distribution of Outer Rim fit concentrations for all 641 779 halos of mass $M_{200c} \geq 6.794 \times 10^{14} \: h^{-1}{ M}_\odot$ at $z=0$. The comparison is with \cite{reed11}, 3501 halos from the Millennium Simulation (black dashed curve; our $6.794 \times 10^{14} \: h^{-1}{ M}_\odot$ with $c_{200c} = 4.367$ corresponds to $M_{\rm vir} = M_{95.4c} = 8.600 \times 10^{14} \: h^{-1}{ M}_\odot$)). For further discussion, see text.}
\end{figure}

As discussed in Section~\ref{sec:ind_halos} above, we find that concentrations within a mass bin are normally distributed. \cite{reed11} and \cite{bhattacharya13} also discuss the Gaussian distribution of their concentration measurements. Figure \ref{fig:conc_gaussian_reed} compares our results in corresponding mass bins. There is very good agreement, in the lowest mass bin, with the results of \cite{bhattacharya13}; the corresponding simulations have very similar mass resolutions. The two higher mass bins of \cite{bhattacharya13} use a simulation with significantly worse mass resolution, which may explain the worse agreement with our results. The agreement with \cite{reed11} is very interesting, considering that the underlying cosmology differs considerably from the one used here. We note that the mass bin here is by no means narrow; our $c$--$M$ relation falls by 25\% from the lowest mass in this mass bin to the highest.

\subsection{Peak Height Parameter $\nu$}
\label{app:nu}
\begin{figure}
	\centering
    \includegraphics{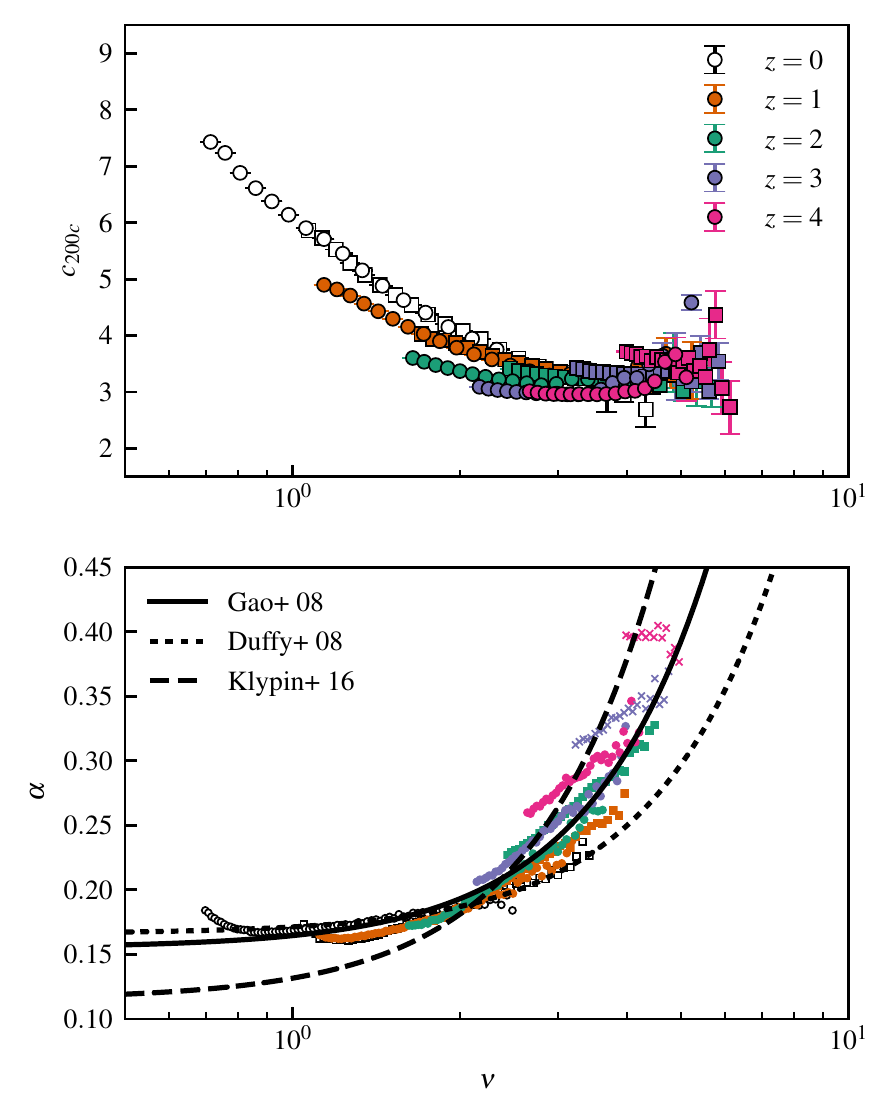}
	\caption{\label{fig:c_nu_alpha_nu} Individual $c_{\rm fit}$ (top panel) and stacked Einasto shape parameter $\alpha$ (bottom panel) as a function of $\nu$ for Q~Continuum (circles) and Outer Rim (squares). As in Figure \ref{fig:alpha_mmstar}, at high $z$, Outer Rim halos are insufficiently resolved ($\times$ symbols). Curves show the fits given by \cite{gao08} (used also by \citealt{dutton14}) and \cite{klypin16}.}
\end{figure}

Efforts to find a redshift-independent expression for profile parameters often (e.g. \citealt{gao08, prada12, dutton14, klypin16}) convert mass to the ``peak height'' parameter $\nu$. Like $M_\star$, $\nu$ is defined from the mean square perturbation $\sigma$ (Equation \ref{eqn:sigma}):
\begin{equation}
\nu(M, z) = \delta_c / \sigma(M, z)
\end{equation}
We again use $\delta_c = 1.686$. Note that $\nu(M_\star, z) = 1$, but $\nu$ is much less sensitive to mass and redshift than $M/M_\star$. In Figure \ref{fig:c_nu_alpha_nu}, we see disagreement between concentrations at different redshifts at low $\nu$. Figure \ref{fig:alpha_mmstar} shows that scaling mass by $M_\star$ does not suggest a redshift-independent scaling for the Einasto shape parameter, but we do not find significant improvement when $\alpha$ is expressed as a function of $\nu$. The $\alpha-\nu$ relation is too flat at high $z$ to fall on the same line as the lower-$z$ relations, as seen in Figure \ref{fig:c_nu_alpha_nu}. However, our results are in reasonable agreement with those of \cite{gao08} for $0 \leq z \leq 3$, $\sigma_8 = 0.9$ (and slightly less so with the Planck cosmology results of \citealt{klypin16}). A discrepancy with the model of \cite{gao08} arises at $z=4$, but the model was developed using results only at $z \leq 3$.

\section{Sources of Error}
Characteristics of the simulation, such as timestep or initial redshift, or of the halo finder, such as the choice of halo center, may affect the shape or amplitude of the $c$--$M$ relation. We check some of the error scenarios in small simulations, and others using the different properties of the Q~Continuum and Outer Rim simulations. We find that the $c$--$M$ relation as obtained here is robust for $z\leq 4$. However, there is more distortion at higher redshifts, where, as expected, the agreement between our two simulations with different mass resolutions deteriorates. Below, we present our tests and comment on the various results obtained.

Unfortunately, none of these tests are able to reproduce the differently shaped $c$--$M$ relations found by other works in Appendix \ref{app:cm_others} above. No two works use the same procedures, so the many remaining methodological differences we are unable to test must contribute to the discrepancies between the resulting $c$--$M$ relations. These include halo identification algorithms, the exclusion of unbound particles from halo profiles, and methods of calculating concentrations, among others.

\subsection{Initial Redshift}
\begin{figure}[htb]
	\centering
    \includegraphics{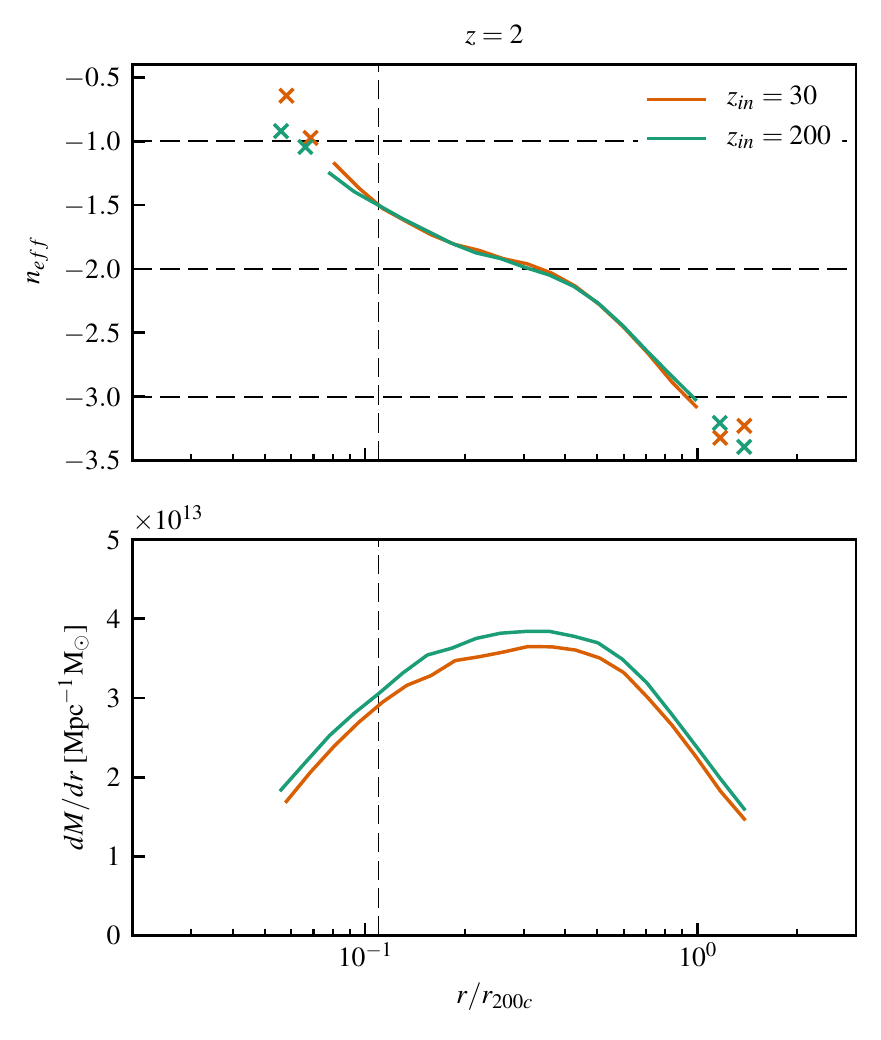}
	\caption{\label{fig:zin} Effect of initial redshift on stacked halo profiles using two simulations with the same realization, but with initial redshifts $z_{\rm in}=200$ and $z_{\rm in}=30$. The figure shows a 1\% stacked profile at $z=2$, with 115 relaxed halos. Mass is lost across the profile (bottom panel) with little effect on $n_\textrm{eff}$ (top); the vertical line shows the innermost radial bin that would be used in fitting this profile. Slopes for the first and last two radial bins ($\times$ symbols) are less reliable than those with four neighboring points to include in the calculation. Mean mass of the stacked $z_{\rm in}=30$ halos is $2.000 \times 10^{13} \: h^{-1}{M}_\odot$; mean mass of their $z_{\rm in} = 200$ pairs is $2.169 \times 10^{13} \: h^{-1}{ M}_\odot$.}
\end{figure}

The large simulations are initialized at $z_{\rm in} = 200$. (In general, the simulations follow the guidelines laid down in \citealt{heitmann10}.) We compare a small simulation (same cosmology with box size $L=1000 \: h^{-1} {\rm Mpc}$ and $m_p=8.56 \times 10^9 \: h^{-1}{M}_\odot$) with this initial redshift to one with a deliberately extreme value of $z_{\rm in} = 30$. While differences in the profiles are visible at $z=2$, the $c$--$M$ relations differ by only a few percent. The two small simulations have the same seed, so halos form at the same locations; two halos with the same center are the same halo and can be directly compared. At $z=2$, 35,551 halos of at least 2000 particles are found in the simulation with $z_{\rm in} = 200$; 28,874 are found in the simulation with $z_{\rm in} = 30$. More halos form in the simulation with higher initial redshift, so we select relaxed halos from $z_{\rm in} = 30$ and pair each with the $z_{\rm in} = 200$ halo whose center is closest, then stack in narrow mass bins. As seen in the stacked profile of Figure~\ref{fig:zin}, the $z_{\rm in}=30$ halos lose mass at all radii (more than 90\% of the $z_{\rm in}=200$ halos have higher mass than their $z_{\rm in}=30$ pairs), while the slope of the profile is unchanged. In particular, note that the radius at which $n_\textrm{eff}=-2$ is identical; the small change in mass, thus $r_{200}$, does produce a slightly higher concentration when $z_{\rm in}=200$. For the example in Figure~\ref{fig:zin}, $c_{\rm fit}$ increases from $3.44$ when $z_{\rm in} = 30$ to $3.49$ when $z_{\rm in} = 200$.

\subsection{Timesteps}
\label{app:timesteps}
\begin{figure}[htb]
	\centering
    \includegraphics{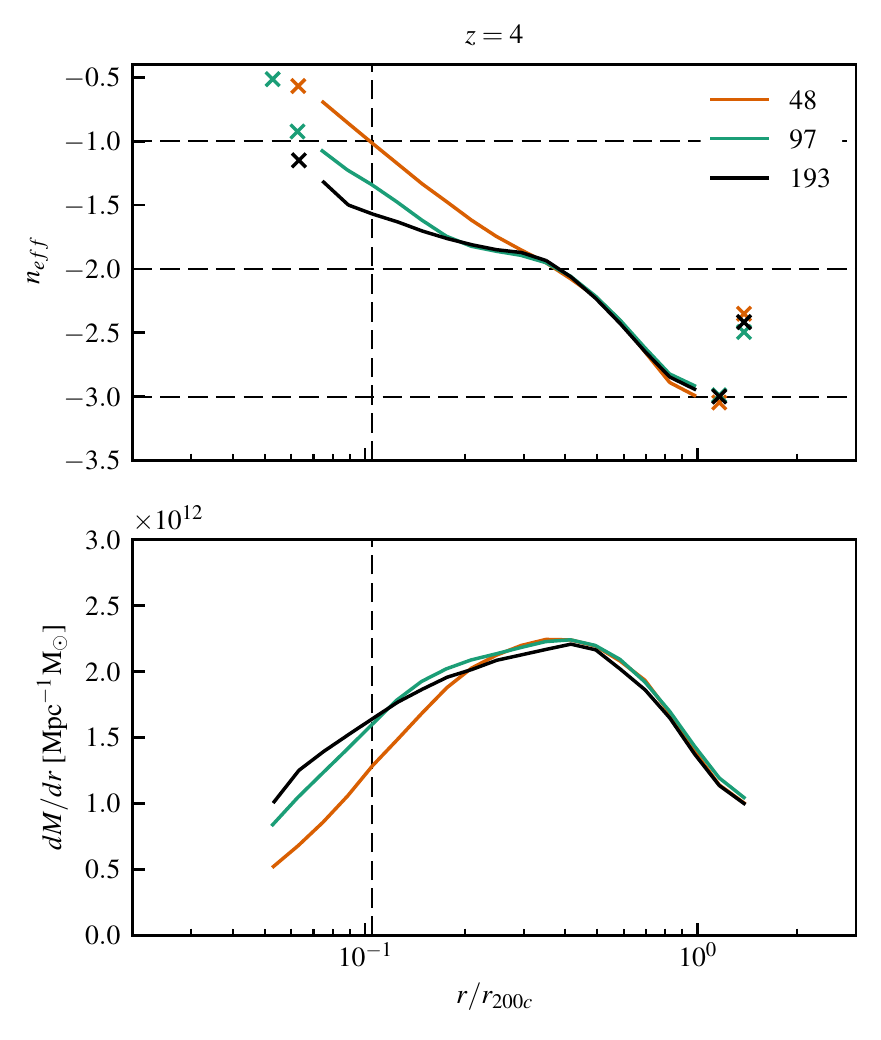}
	\caption{\label{fig:timestep} Effect of number of timesteps on stacked halo profiles: a 5\% stacked profile at $z=4$, 116 relaxed halos. The inner profile converges with more timesteps; the vertical line shows the innermost radial bin that would be used in fitting this profile. Slopes for the first and last two radial bins ($\times$ symbols) are less reliable than those with four neighboring points to include in the calculation. The mean mass of the stacked  halos with 48 timesteps is $2.983 \times 10^{11} \: h^{-1}{ M}_\odot$; mean mass of their pairs is $3.069 \times 10^{11} \: h^{-1}{ M}_\odot$ with 97 timesteps, and $3.011 \times 10^{11} \: h^{-1}{ M}_\odot$ with 193 timesteps.}
\end{figure}

In this test, we run smaller simulations (same cosmology with box size $L=115.375 \: h^{-1}\:{\rm Mpc}$ and same mass resolution as Q~Continuum, $m_p=1.05 \times 10^8 \: h^{-1}{ M}_\odot$) with half, twice, and the same number of timesteps as for the larger simulations. Again, all three test simulations use the same seed, so the profiles of paired halos can be compared. At $z=4$, 2782 halos of at least 2000 particles are found in the simulation with the standard number of timesteps; 2625 are found in the simulation with halved timesteps, and 2705 when timesteps are doubled. The fewest halos form with halved timesteps, so relaxed halos are selected from that simulation and paired with the closest halos in the others. As shown in Figure \ref{fig:timestep}, the profile at large radius is unchanged, but the inner profile requires more timesteps to converge. Note that all three profiles cross $n_\textrm{eff}=-2$ and peak at the same point, indicating no change in scale radius. The shape of the $c$--$M$ relation is essentially unaffected; at $z=3$, doubling the number of timesteps shifts the $c$--$M$ relation uniformly up by less than 5\% for both fit and accumulated mass methods, while halving timesteps shifts it down by about 8\%.

\subsection{Halo Centering}
We identify the center of a halo as the most bound particle (MBP), locating the local minimum of the potential. Other methods select the most connected particle (MCP, the particle with the greatest number of FOF neighbors), or use a histogram method to find the maximum density. We compare these three methods for a test simulation at $z=0$. For relaxed halos, where the MBP and MCP centers can differ by $60\:h^{-1}\:{\rm kpc}$, the concentrations vary by less than 5\%. The distance between centers is greater for unrelaxed halos -- an average of $160\:h^{-1}\:{\rm kpc}$. At low masses, their fit concentrations differ as well; MCP and histogram fit concentrations are up to 20\% greater than MBP for halos of 2000 particles, $M_{200c} = 2 \times 10^{13} \: h^{-1}{M}_\odot$, while the difference returns to less than 5\% at high masses.

\subsection{Inner Profile}
\label{app:inner_profile}
\begin{figure}[htb]
	\centering
	\includegraphics{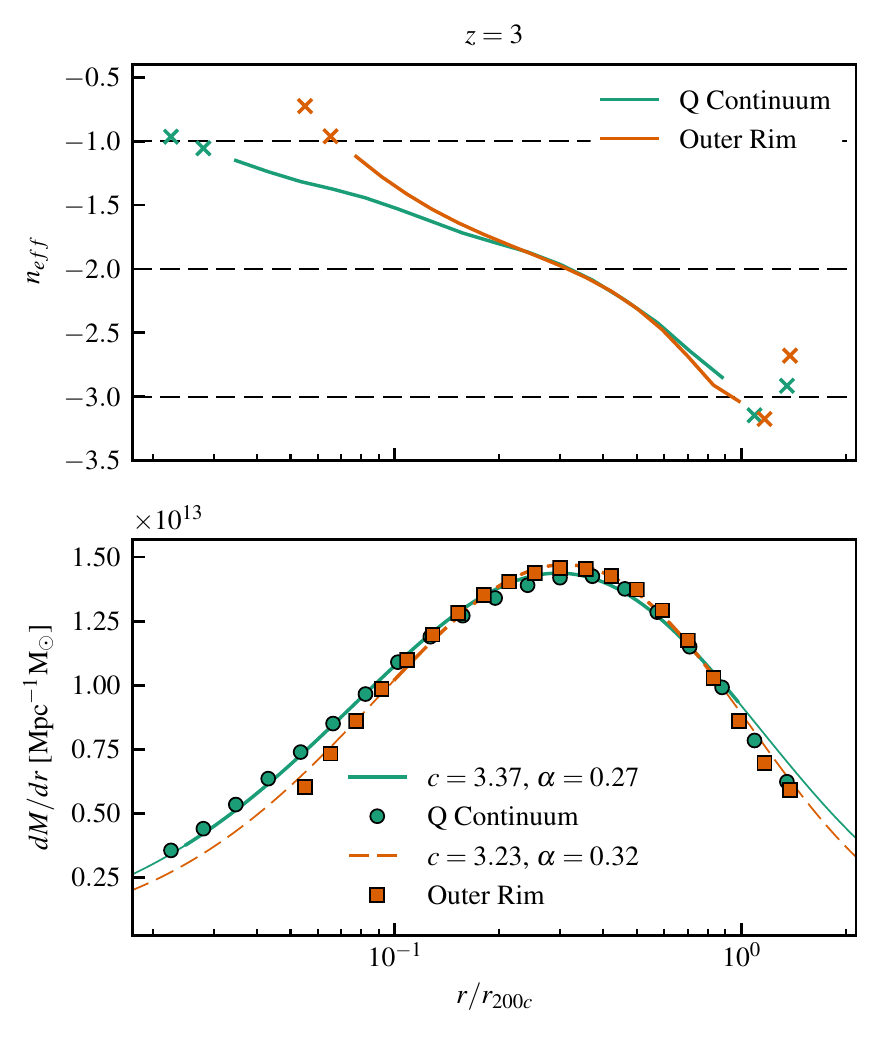}
	\caption{\label{fig:inner_profile} Einasto fit to 9556 stacked Outer Rim and 254 Q Continuum profiles at $z=3$, $M_{200c} = 5 \times 10^{12} \: h^{-1}{ M}_\odot \pm 1\%$. Slopes for the first and last two radial bins ($\times$ symbols, top panel) are less reliable than those with four neighboring points to include in the calculation. Einasto curves are thin in the regions not included in the fit ($r > r_{200}$ or fewer than 100 particles enclosed). In this example, the change in the inner profile causes a 20\% discrepancy in fit shape parameter $\alpha$ (see the gap between Outer Rim and Q~Continuum high-$z$ shape parameters in Fig. \ref{fig:alpha_mmstar}), but a less than 5\% difference in fit concentration. }
\end{figure}

As seen in Figures~\ref{fig:stack1} and \ref{fig:timestep}, differences from the `true' profile are more pronounced in the inner profile than at large radius. Error in the inner profile may have little effect on measurements of NFW concentration, but the Einasto profiles are more sensitive to profile shape; a small change in the inner profile produces a larger change in the shape parameter $\alpha$ than in the concentration, as seen in Figures~\ref{fig:alpha_mmstar} and \ref{fig:inner_profile}. Inner-profile discrepancies between Outer Rim and Q~Continuum arise only at high redshift; where measurements of the shape parameter are inconsistent, the higher-resolution Q~Continuum profiles are closer to the `true' shape. The convergence in the inner profile is consistent with the considerations presented in \cite{power03}.

\subsection{Peak-finding Method}
\label{app:peak_method}
We present $c_{\rm peak}$ only for lower-concentration halos; this is because of a systematic error that can arise due to insufficient mass resolution. Figure~\ref{fig:particle_count} shows the artificially high $c_{\rm peak}$ measurements found at lower masses at $z=0$: when concentration is high, $r_s$ is found at smaller radii, where error in particle counts per radial bin may be substantial. For a low-mass Outer Rim halo at $z=0$ with mass $4 \times 10^{12} \: h^{-1}{M}_\odot$ and concentration $c \sim 6$, $r_s$ is found around $0.07 \: h^{-1}\textrm{Mpc}$. A single radial bin at this radius will contain about 100 particles; even with three-point smoothing, the peak is noisy. The other two methods are less susceptible to this error: accumulated mass sums over neighboring bins, and profile fit takes the Poisson error in each bin into account. At high masses and low concentrations, however, the relevant radial bins contain 1--2 orders of magnitude more particles; the peak is better defined and the peak-finding method finds concentrations more similar to the results of the other methods. 

\subsection{Minimum Particle Count}
\label{sec:min_particle_count}
\begin{figure*}
	\centering
    \includegraphics{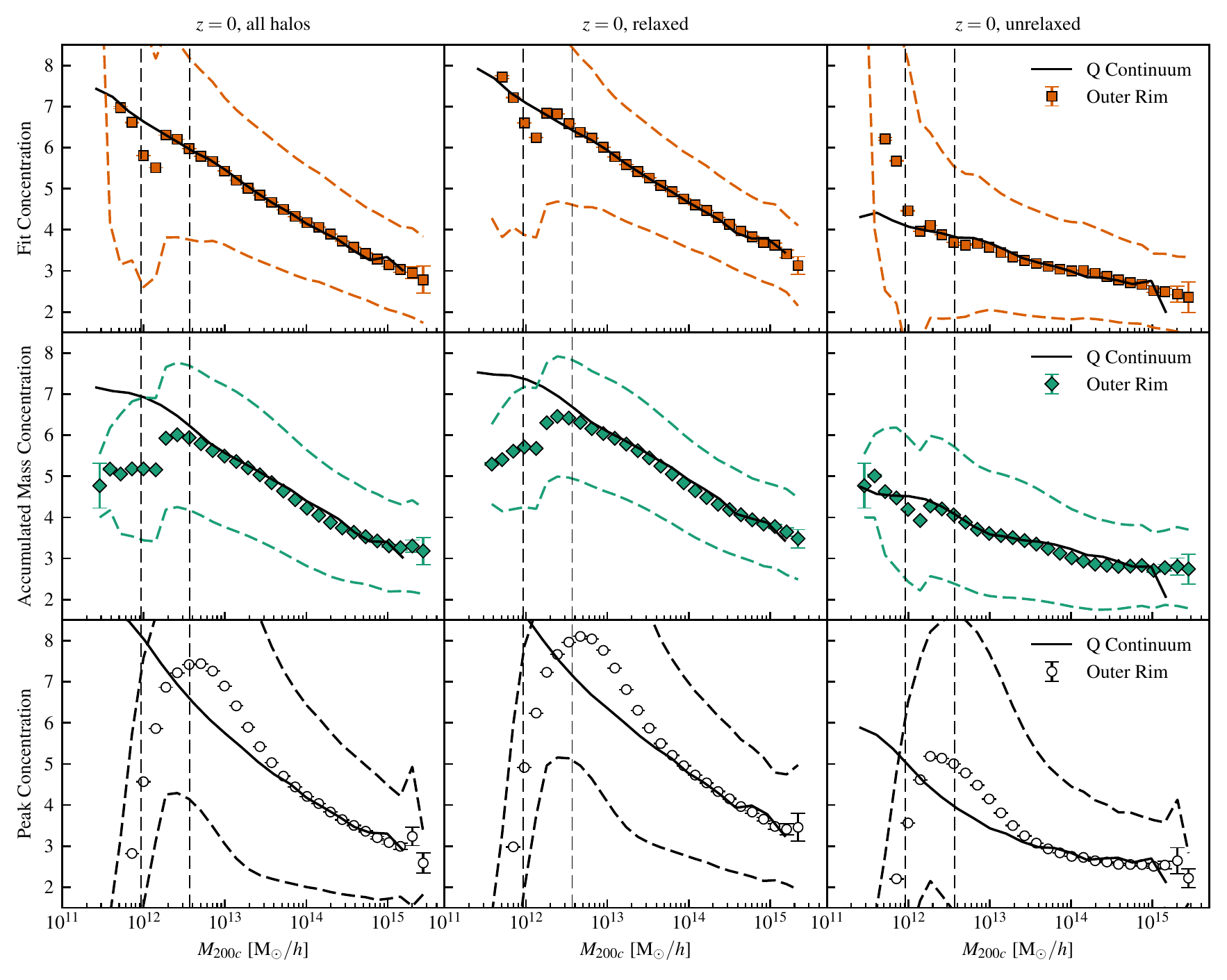}
	\caption{\label{fig:particle_count} At $z=0$, all three methods agree across simulations when particle count is high, but not on halos containing fewer particles. A Q~Continuum halo of mass $M_{200c} = 2.1 \times 10^{11} \: h^{-1}{ M}_\odot$ contains 2000 particles, while an Outer Rim halo of the same mass contains only 114. Dashed vertical lines show mass corresponding to 500 and 2000 Outer Rim particles; the minimum particle count of Q Continuum halos shown is 2000. Mean Outer Rim fit and accumulated mass concentrations (points) diverge from Q Continuum concentrations (lines) below 2000 particles; the peak method is only reliable on halos with an order of magnitude more particles. Dashed curves show 1$\sigma$ intrinsic variance of Outer Rim halo concentrations.}
\end{figure*}

We present concentrations only for halos of at least 2000 particles, as in~\cite{bhattacharya13}. \cite{neto07} discuss conservative minimum particle counts of $10^3 - 10^4$ to ensure high-quality fits and agreement between different mass resolutions. \cite{povedaruiz16}, finding statistical bias in concentration measurements of halos resolved with hundreds of particles, advocate a minimum particle count of approximately 4000 particles. Because the mass resolutions of Outer Rim and Q Continuum differ by a factor of 20, the two simulations can be used to check for convergence at fixed mass. (The concentrations of 100-particle Outer Rim halos, for example, can be compared to 2000-particle Q Continuum halos of the same mass.) As seen in Figure~\ref{fig:particle_count}, 2000 particles is sufficient for fit and accumulated mass methods to agree, but, as discussed above, the peak method can be used safely only on larger, less concentrated halos. 

\end{document}